\documentclass[reqno,11pt]{amsart}
\usepackage{graphicx}
\usepackage{slashed}
\usepackage{amscd}
\usepackage{tensor}
\usepackage{amssymb}
\usepackage{esint}
\usepackage{enumitem}
\usepackage{accents}
\usepackage[dvipsnames]{xcolor}
\usepackage[utf8]{inputenc}	
\usepackage{pst-grad} 
\usepackage{pst-plot} 
\usepackage[mathscr]{eucal}
\numberwithin{equation}{section}
\allowdisplaybreaks[1]

\title[Conserved Surface Layer Integrals]{A Class of Conserved Surface Layer Integrals \\
for Causal Variational Principles}

\author[F.\ Finster]{Felix Finster}
\address{Fakult\"at f\"ur Mathematik \\ Universit\"at Regensburg \\ D-93040 Regensburg \\ Germany}
\email{finster@ur.de}
\author[J.\ Kleiner]{Johannes Kleiner \\ \\ December 2018}
\address{Institut f\"ur Theoretische Physik \\ Leibniz Universit\"at \\ D-30167 Hannover \\ Germany}
\email{johannes.kleiner@itp.uni-hannover.de}

\newtheorem{Def}{Definition}[section]
\newtheorem{Thm}[Def]{Theorem}
\newtheorem{Prp}[Def]{Proposition}
\newtheorem{Lemma}[Def]{Lemma}

\newtheorem{Corollary}[Def]{Corollary}
\newtheorem{Example}[Def]{Example}

\newcommand{\Thanks}{\vspace*{.5em} \noindent \thanks}
\newcommand{\beq}{\begin{equation}}
\newcommand{\eeq}{\end{equation}}
\newcommand{\Proof}{\begin{proof}}
\newcommand{\QED}{\end{proof} \noindent}
\newcommand{\QEDrem}{\ \hfill $\Diamond$}

\newcommand{\la}{\langle}
\newcommand{\ra}{\rangle}

\newcommand{\R}{\mathbb{R}}
\newcommand{\1}{\mbox{\rm 1 \hspace{-1.05 em} 1}}
\newcommand{\Z}{\mathbb{Z}}
\newcommand{\N}{\mathbb{N}}

\renewcommand{\L}{{\mathcal{L}}}
\newcommand{\Sact}{{\mathcal{S}}}

\DeclareMathOperator{\supp}{supp}

\newcommand{\F}{{\mathscr{F}}}

\DeclareFontFamily{OT1}{rsfso}{}
\DeclareFontShape{OT1}{rsfso}{m}{n}{ <-7> rsfso5 <7-10> rsfso7 <10-> rsfso10}{}
\DeclareMathAlphabet{\mycal}{OT1}{rsfso}{m}{n}

\newcommand{\Jdiff}{\mathfrak{J}^\text{\rm{\tiny{diff}}}}
\newcommand{\Jtest}{\mathfrak{J}^\text{\rm{\tiny{test}}}}
\newcommand{\Jlin}{\mathfrak{J}^\text{\rm{\tiny{lin}}}}
\newcommand{\Gdiff}{\Gamma^\text{\rm{\tiny{diff}}}}
\newcommand{\Gtest}{\Gamma^\text{\rm{\tiny{test}}}}

\newcommand{\Ctest}{C^\text{\rm{\tiny{test}}}}
\renewcommand{\u}{\mathfrak{u}}
\renewcommand{\v}{\mathfrak{v}}
\newcommand{\w}{\mathfrak{w}}
\newcommand{\J}{\mathfrak{J}}
\newcommand{\UL}{{\mathcal{Z}}}		
\newcommand{\Fock}{{\mathcal{F}}}
\newcommand{\bitem}{\begin{itemize}[leftmargin=2.5em]}
\newcommand{\eitem}{\end{itemize}}
\newcommand{\itemD}{\item[{\raisebox{0.125em}{\tiny $\blacktriangleright$}}]}
\newcommand{\Pert}{{\mathscr{P}}}

\begin{document}

\maketitle

\begin{abstract}
In the theory of causal fermion systems, the physical equations
are obtained as the Euler-Lagrange equations of
a causal variational principle.
Studying families of critical measures of causal variational principles,
a class of conserved surface layer integrals is found and analyzed.
\end{abstract}

\tableofcontents

\section{Introduction}
In the theory of causal fermion systems, which is a recent approach to fundamental physics,
the physical equations are obtained by minimizing the so-called causal action
under variations of a measure.
Causal variational principles were introduced in~\cite{continuum}
as a generalization of this causal action principle (see~\cite{cfs} or the survey article~\cite{dice2014}).
In the meantime, causal variational principles have been studied in various situations,
and the mathematical setup has been further extended (see~\cite{jet} for a recent introduction).
In general terms, in a {\em{causal variational principle}} one minimizes an {\em{action}}~$\Sact$ of the form
\beq \label{Sact} 
\Sact (\rho) = \int_\F d\rho(x) \int_\F d\rho(y)\: \L(x,y) 
\eeq
under variations of the measure~$\rho$, keeping the total volume~$\rho(\F)$ fixed
(for more details see Section~\ref{secprelim} below).
The support of the measure is referred to as
\[ \text{space-time} \qquad M:= \supp \rho \:. \]
In this setting, the usual integrals over hypersurfaces in space-time are undefined.
Instead, one considers so-called {\em{surface layer integrals}}, being double integrals of the form
\beq \label{IntrOSI}
\int_\Omega d\rho(x) \int_{M \setminus \Omega} d\rho(y) \:(\cdots)\: \L(x,y) \:,
\eeq
where~$\Omega$ is a subset of~$M$ and~$(\cdots)$ stands for a differential operator
acting on the Lagrangian. The structure of such surface layer integrals can be understood most easily 
in the special situation that the Lagrangian is of short range
in the sense that~$\L(x,y)$ vanishes unless~$x$ and~$y$ are close together.
In this situation, we get a contribution to the double integral~\eqref{IntrOSI} only
if both~$x$ and~$y$ are close to the boundary~$\partial \Omega$.
With this in mind, surface layer integrals can be understood as an adaptation
of surface integrals to the setting of causal variational principles
(for a more detailed explanation see~\cite[Section~2.3]{noether}).

Surface layer integrals were first introduced in~\cite{noether} in order to
formulate Noether-like theorems for causal variational principles.
In particular, it was shown that there is a conserved
surface layer integral which generalizes the Dirac current
in relativistic quantum mechanics (see~\cite[Section~5]{noether}).
More recently, in~\cite{jet} another conserved
surface layer integral was discovered which gives rise to a symplectic form on the
solutions of the linearized field equations (see~\cite[Sections~3.3 and~4.3]{jet}).
The present paper is devoted to a systematic study of conserved surface layer integrals.
We find a class of conserved surface layer integrals~$I^\Omega_m$ parametrized
by a parameter~$m=1,2,\ldots$ (see Theorem~\ref{thmosirho}).
These surface layer integrals are obtained by considering families of
critical points of the causal variational principle.
They can be rewritten as multilinear functionals on the space~$ \Jlin \cap \Jtest$
of solutions of the linearized field equations (see Theorem~\ref{thmlin}).
In the case~$m=1$, the resulting surface layer integrals have the same structure as
those in the Noether-like theorems in~\cite{noether} (see Corollary~\ref{ex0}).
In the case~$m=2$, anti-symmetrizing the bilinear functional in its two arguments~$\u, \v \in \Jlin \cap \Jtest$
gives the symplectic form  (see Corollary~\ref{exsymp}). Symmetrizing in these arguments, on the other hand,
gives rise to a new conserved surface layer integral
of the following form (see also Corollary~\ref{exsprod}):
\begin{Thm} {\bf{(symmetric bilinear form on jets)}} \label{thmintro}
Given a compact set~$\Omega \subset M$, if the regularity
assumptions~(r1) and~(r2) on page~\pageref{r1} are satisfied, then the bilinear form
\begin{align}
(.,.)_\Omega &\::\: (\Jlin \cap \Jtest) \times (\Jlin \cap \Jtest) \rightarrow \R \notag \\
\begin{split}
(\u,\v)_\Omega &= \int_\Omega d\rho(x) \int_{M \setminus \Omega} d\rho(y)\; \\
&\!\!\!\!\!\!\!\!\!\!\!\!\!\!\times
\big( \nabla_{1, \u} \nabla_{1, \v} + 2 \,\nabla_{1, S \Delta_2[ \u, \v]}
- \nabla_{2, \u} \nabla_{2, \v} - 2 \,\nabla_{2, S \Delta_2[ \u, \v]}
\big) \L(x,y) \label{osiform}
\end{split}
\end{align}
satisfies the relation
\beq \label{osirel}
(\u,\v)_\Omega = \int_\Omega \Big(  \nabla_{1,\u} \nabla_{1,\v} \: \frac{\nu}{2} + 
\nabla_{S \Delta_2[\u,\v]} \: \nu \Big) \:d\rho(x) \:.
\eeq
\end{Thm} \noindent
Here the jets $\u$ and $\v$ are pairs of a real-valued function on~$M$ and a vector field on~$\F$ along~$M$;
more precisely,
\[ \u=(a,u),\:  \v=(b,v) \;\in\; C^\infty(M, \R) \oplus C^\infty(M, T\F) \]
(where by~$C^\infty(M,.)$ we denote functions which have a smooth extension to~$\F$;
for details see the explanation after~\eqref{SmoothVecFields} below). Moreover, $\nabla_\u$ denotes a
\textit{jet derivative}, which is a linear combination of multiplication and directional derivative,
\beq \label{Djet}
\nabla_{\u} \ell(x) := a(x)\, \ell(x) + \big(D_u \ell \big)(x) \:.
\eeq
The indices in $\nabla_{1,\u}$ or $\nabla_{2,\u}$ indicate on which argument of the Lagrangian the 
jet derivative acts. We always use the convention that these ``partial jet derivatives'' do not act on
jets contained in other derivatives, so that for example for a constant $\nu$ (for
details see Section~\ref{secprelim}),
\[ \nabla_{1,\u} \nabla_{1,\v} \: \nu = a(x) \,b(x) \: \nu \:. \]
The linearized field equations for a jet~$\v$ read (for details see Section~\ref{LinSol})
\[ \nabla_\u \bigg( \int_{M} \big( \nabla_{1, \v} + \nabla_{2, \v} \big) \L(x,y)\: d\rho(y)
-\nabla_\v \: \frac{\nu}{2} \bigg) = 0 \:, \]
to be satisfied on~$M$ for all test jets~$\u$ in the jet space~$\Jtest$
(see~\eqref{JtestDef} and Definition~\ref{deflinear} or~\cite[Section~4.2]{jet}).
The space of solutions of the linearized field equations is denoted by~$\Jlin$.
Next, $\Delta_2$ is the quadratic correction to the field equations~\eqref{lin},
and~$S$ is a Green's operator (see Definition~\ref{Sdefine};
a basic introduction will be given in Section~\ref{secprelim}).
The surface layer integral~\eqref{osiform} corresponds to a {\em{surface term}},
whereas the right side of~\eqref{osirel} involves an integral over~$\Omega$
and is therefore a {\em{volume term}}.
If the jets~$\u, \v$ and~$S \Delta_2[ \u, \v]$ have no scalar components,
then the volume term vanishes, thereby giving a conservation law for the surface layer integral
(see Corollary~\ref{exsprod}).

This paper is structured as follows. In Section~\ref{secprelim} we provide the necessary preliminaries and
collect all the assumptions which enter our constructions.
Section~\ref{ClassSLI} contains our main results. In Section~\ref{ConsFamSol} we prove a
general theorem which yields conserved quantities and illustrate it in the example
of symmetry transformations.
This theorem involves a two-parameter family of solutions of the 
weak EL equations.
In Section~\ref{LinSol} we compute the surface layer integrals
for a two-parameter family constructed from~$m$ solutions of the
linearized field equations (see Theorem~\ref{thmlin}),
using perturbative methods as developed in~\cite{perturb}.
In Section~\ref{DepGreensOp} we discuss the dependence of the bilinear form~\eqref{osiform}
on the choice of the Green's operator~$S$. This freedom changes the bilinear form only modulo a
conserved surface layer integral~$I^\Omega_1$ (see Example~\ref{ex1green}). 
More generally, modifying the Green's operators in~$I^\Omega_m$ changes the surface layer integrals only by
multiples of~$I^\Omega_{l}$ with~$l<m$ (see Theorem~\ref{thmgreen}).
In Section~\ref{seclattice} our constructions are illustrated by computing the
surface layer integrals in the example of the lattice system
introduced in~\cite[Section~5]{jet}.
Finally, in Section~\ref{secoutlook} we explain the physical implications
of our results and give a brief outlook.

\section{Preliminaries} \label{secprelim}
We consider the lower semi-continuous setting as introduced in~\cite[Section~2]{jet}.
Thus let~$\F$ be a (possibly non-compact) smooth manifold
and~$\rho$ a {\em{Radon measure}} on~$\F$ (i.e.\ a regular Borel measure
with~$\rho(K)<\infty$ for any compact~$K \subset \F$, where
by a measure we always mean a {\em{positive}} measure; for an introduction see for
example~\cite{halmosmt} or~\cite{bogachev}), referred to as the {\em universal measure}.
Moreover, we are given a non-negative function~$\L : \F \times \F \rightarrow \R^+_0$
(the {\em{Lagrangian}}) with the following properties:
\begin{itemize}[leftmargin=2em]
\item[(i)] $\L$ is symmetric: $\L(x,y) = \L(y,x)$ for all~$x,y \in \F$.\label{Cond1}
\item[(ii)] $\L$ is lower semi-continuous,  \label{Cond2}
i.e.\ for all sequences~$x_n \rightarrow x$ and~$y_{n'} \rightarrow y$,
\[ \L(x,y) \leq \liminf_{n,n' \rightarrow \infty} \L(x_n, y_{n'})\:. \]
\end{itemize}
We assume that~$\rho$ satisfies the following technical assumption:
\begin{itemize}[leftmargin=2em]
\item[(iii)] The function~$\L(x,.)$ is $\rho$-integrable for all~$x \in \F$, giving
a lower semi-continuous and bounded function on~$\F$. \label{Cond4}
\end{itemize}
If the total volume~$\rho(\F)$ is finite, the {\em{causal variational principle}} is to minimize the
action~\eqref{Sact} under variations of the measure~$\rho$
(which do not need to satisfy~(iii)), keeping the total volume~$\rho(\F)$ fixed
({\em{volume constraint}}).
If~$\rho(\F)$ is infinite, however, it is not obvious how to implement the volume constraint,
making it necessary to proceed as follows:
Let~$\tilde{\rho}$ be another Borel measure on~$\F$ with the properties
\beq \label{totvol}
\big| \tilde{\rho} - \rho \big|(\F) < \infty \qquad \text{and} \qquad
\big( \tilde{\rho} - \rho \big) (\F) = 0
\eeq
(where~$|.|$ denotes the total variation of a measure;
see~\cite[\S28]{halmosmt} or~\cite[Section~6.1]{rudin}).
Then the difference of the actions as given by
\beq \label{integrals}
\begin{split}
\big( &\Sact(\tilde{\rho}) - \Sact(\rho) \big) = \int_\F d(\tilde{\rho} - \rho)(x) \int_\F d\rho(y)\: \L(x,y) \\
&\quad + \int_\F d\rho(x) \int_\F d(\tilde{\rho} - \rho)(y)\: \L(x,y) 
+ \int_\F d(\tilde{\rho} - \rho)(x) \int_\F d(\tilde{\rho} - \rho)(y)\: \L(x,y)
\end{split}
\eeq
is well-defined (for details see~\cite[Lemma~2.1]{jet}). 
The measure~$\rho$ is said to be a {\em{minimizer}} of the causal action
if the difference~\eqref{integrals}
is non-negative for all~$\tilde{\rho}$ satisfying~\eqref{totvol},
\[ \big( \Sact(\tilde{\rho}) - \Sact(\rho) \big) \geq 0 \:. \]

We now state the Euler-Lagrange (EL) equations as derived
in~\cite[Lemma~2.3]{jet} (by adapting~\cite[Lemma~3.4]{support} to the non-compact setting).
\begin{Lemma} (The Euler-Lagrange equations) \label{lemmaEL}
Let~$\rho$ be a minimizer of the causal action. Then
for a suitable value of the real parameter~$\nu$, the function~$\ell$ defined by
\beq \label{elldef}
\ell(x) := \int_\F \L(x,y)\: d\rho(y) - \frac{\nu}{2} \::\: \F \rightarrow \R
\eeq
satisfies the equation
\begin{align}\label{ELstrong}
\ell|_{\supp \rho} \equiv \inf_\F \ell = 0 \: .
\end{align}
\end{Lemma} \noindent
We remark that~$\nu$ can be understood as the Lagrange multiplier describing the volume constraint;
see~\cite[\S1.4.1]{cfs}.

The EL equations are analyzed most conveniently in the so-called \textit{jet formalism},
which we now review (for more details see~\cite{jet}).
By~$C^\infty(M, \R)$ we denote all real-valued functions on~$M$ which have a smooth
extension to~$\F$. Likewise, by
\beq\label{SmoothVecFields}
\Gamma = C^\infty(M, T\F)
\eeq
we denote the smooth vector fields on~$\F$ along~$M$
(thus every $u \in \Gamma$ is the restriction of a smooth vector field on~$\F$ to~$M$).
We define the jet space on~$M$ as the vector space
\beq \label{Jdef}
\J := \big\{ \u = (a,u) \text{ with } a \in C^\infty(M, \R) \text{ and } u \in \Gamma \big\} \:.
\eeq
Moreover, we let~$\Gdiff$ be those vector fields for which the
directional derivative of the function~$\ell$ exists,
\[ \Gdiff = \big\{ u \in C^\infty(M, T\F) \;\big|\; \text{$D_{u} \ell(x)$ exists for all~$x \in M$} \big\} \:. \]
We introduce the space of \textit{differentiable one-jets} by
\beq
\label{Jdiff}
\Jdiff := C^\infty(M, \R) \oplus \Gdiff \;\subset\; \J
\eeq
and choose a linear subspace~$\Jtest$ of the form
\beq \label{JtestDef}
\Jtest = \Ctest(M, \R) \oplus \Gtest \;\subset\; \Jdiff \:,
\eeq
where the vector space of the scalar component~$\Ctest(M, \R)$ is assumed to be nowhere trivial in the sense that
\beq \label{Cnontriv}
\text{for all~$x \in M$ there is~$a \in \Ctest(M, \R)$ with~$a(x) \neq 0$}\:.
\eeq
For a jet~$\u = (a, u) \in \Jdiff$ we define~$\nabla_\u$ as the linear combination of
scalar multiplication and directional derivative~\eqref{Djet}.
Then the EL equations~\eqref{ELstrong} imply the so-called {\em{weak EL equations}} (for details see~\cite[Section~4.1]{jet})
\beq \label{ELtest}
\nabla_{\u} \ell|_M = 0 \qquad \text{for all~$\u \in \Jtest$}\:.
\eeq
The purpose of introducing~$\Jtest$ is that it gives the freedom to restrict attention to the portion of
information in the EL equations which is relevant for the particular application in mind.
For example, if one is interested only in the macroscopic dynamics, one can choose~$\Jtest$
to be composed of jets which disregard 
microscopic fluctuations of~$\ell$.

We finally point out that the weak EL equations~\eqref{ELtest}
do not hold only for minimizers, but also for critical points of
the causal action. With this in mind, all methods and results of this paper do not apply only to
minimizers, but more generally to {\em{critical points}} of the causal variational principle.

\section{A Class of Conserved Surface Layer Integrals}\label{ClassSLI}
\subsection{Conservation Laws for Families of Solutions}\label{ConsFamSol}
In this section we derive a class of conservation laws for families of solutions of the
weak EL equations. We let~$\tilde{\rho}_{s,t}$ with~$s, t \in (-\delta, \delta)$ be a two-parameter family of
universal measures of the form
\beq \label{rhost}
\tilde{\rho}_{s,t} = (F_{s,t})_* \big( f_{s,t} \, \rho \big) \:,
\eeq
where~$f_{s,t}$ and~$F_{s,t}$ are smooth functions on~$M$,
\beq \label{fFdef}
f \in C^\infty\big((-\delta, \delta)^2 \times M,\R^+ \big) \qquad \text{and} \qquad
F \in C^\infty\big((-\delta, \delta)^2 \times M, \F \big)
\eeq
(where smooth on~$M$ again means that there exists a smooth extension to~$\F$),
which are trivial at~$s=t=0$,
\beq \label{fFinit}
f_{0,0} \equiv 1\:,\qquad F_{0,0} = \1 \: .
\eeq
(The star in~\eqref{rhost} denotes the push-forward measure, defined 
for a subset~$U \subset \F$ and a measure~$\mu$ on~$\F$ by $((F_{s,t})_* \mu)(U)
= \mu(F_{s,t}^{-1}(U))$; see for example~\cite[Section~3.6]{bogachev}.)
Moreover, we assume that every measure $\tilde\rho_{s,t}$ satisfies the weak EL equations~\eqref{ELtest}.
Since the jets~\eqref{JtestDef} used for testing are
defined on the support of the corresponding measure, the variation of~$\rho$
also entails a variation of the test space $\Jtest$.
We denote the test space corresponding to the measure $\tilde\rho_{s,t}$ by $\tilde{\J}^\text{\tiny{test}}_{s,t}$, i.e.\
\[ \tilde{\J}^\text{\tiny{test}}_{s,t} \subset \big\{ \u = (a,u) \text{ with } a \in C^\infty(\tilde{M}_{s,t}, \R) \text{ and }
u \in C^\infty(\tilde{M}_{s,t}, T\F) \big\} \:, \]
where~$\tilde{M}_{s,t}:= \supp \tilde{\rho}_{s,t}$ is the support of the varied measure.
We write the weak EL equations~\eqref{ELtest} for the measure~$\tilde\rho_{s,t}$ as
\beq \label{ELtest0}
\nabla_{1,\tilde{\u}(F_{s,t}(x))}  \bigg( \int_M \L\big(F_{s,t}(x), F_{s,t}(y) \big)\: f_{s,t}(y) \, d\rho(y) - \frac{\nu}{2} \bigg) = 0 \qquad \text{for all~$\tilde{\u} \in \tilde{\J}^\text{\rm{\tiny{test}}}_{s,t}$} \:,
\eeq
to be evaluated pointwise for all~$x \in M$. Here the notation $\nabla_{1,\tilde{\u}}$
clarifies that the derivative acts on the first argument of the Lagrangian.
On the constant~$\nu/2$ it acts by multiplication with the scalar component,
\[ \nabla_{1,\tilde{\u}(F_{s,t}(x))} \,\nu = \nabla_{\tilde{\u}(F_{s,t}(x))} \,\nu
= a\big(F_{s,t}(x)\big)\, \nu \:, \]
where we again denote the components by~$\u=(a,u)$.
Without loss of generality we can keep $\nu$ fixed when varying the measure.
Note that, being defined on~$\tilde{M}$, the jet~$\tilde{\u}$ can be evaluated at~$x \in M$ 
only after composing it with~$F_{s,t}$.
In order to rewrite this equation in a way where~$x$ and~$y$ are treated in a more symmetric way,
we multiply~\eqref{ELtest0} by the function~$f_{s,t}(x)$. 
Again using our convention that the derivative~$\nabla_{1,\tilde{\u}}$ acts only on the first
argument of the Lagrangian $\L$, the function $f_{s,t}(x)$ commutes with $\nabla_{1,\tilde{\u}}$.
We thus obtain
\beq \label{ELtest_st}
\nabla_{1,\tilde{\u}(F_{s,t}(x))} \bigg( \int_M L\big(x_{s,t}, y_{s,t} \big) \, d\rho(y) - 
\frac{\nu}{2}\:f_{s,t}(x) \bigg) = 0 \qquad \text{for all~$\u \in \tilde{\J}^\text{\rm{\tiny{test}}}_{s,t}$} \:,
\eeq
where we introduced the notation
\beq \label{Ldef}
L\big(x_{s,t}, y_{s,t} \big) := f_{s,t}(x)\: \L\big( F_{s,t}(x), F_{s,t}(y) \big)\: f_{s,t}(y) \:.
\eeq
We note that in~\cite[Section~4]{jet} a slightly different method was used where
we tested with jets on~$M$. The method here is more general
and has the advantage that, as will be explained at the beginning of Section~\ref{LinSol},
it fits together better with the
description in terms of linearized solutions in Section~\ref{LinSol}
as well as with the perturbative description in Section~\ref{secperturb}.

In the next theorem we derive a family of integral identities,
parametrized by an integer~$m=1,2,\ldots$ which describes the order of differentiation of the
Lagrangian. For technical simplicity, we impose the following regularity assumptions:
\begin{itemize}[leftmargin=3em]
\item[\rm{(r1)}] For all~$p,q \in \{0,1\}$ with~$p+q=1$ and all~$x \in M$,
the following partial derivatives exist, are $\rho$-integrable and may be interchanged with the integral:\label{r1}
\begin{align}\begin{split}\label{r1Eq}
\int_M &\partial_{s'}^p \partial_s^q \partial_t^{m-1} \, L\big( x_{s+s',t}, y_{s,t} \big) \Big|_{s'=s=t=0} \:d\rho(y) \\
&= \partial_{s'}^p \partial_s^q \partial_t^{m-1}  \int_M L\big( x_{s+s',t}, y_{s,t} \big) \:d\rho(y) \Big|_{s'=s=t=0} \:. 
\end{split}\end{align}
\item[\rm{(r2)}] The jet~$\tilde{\u}(t)$ defined by\label{r2}
\beq\label{r2Eq}
\tilde{\u}(t) = \partial_s \big( f_{s,t}, F_{s,t} \big) \big|_{s=0}
\qquad \text{is in~$\tilde{\J}^\text{\rm{\tiny{test}}}_{0,t}$ for all~$t \in (-\delta, \delta)$}\:. 
\eeq
\end{itemize}

\begin{Thm} \label{thmosirho}
Let~$m \in \N$. Assume that the functions~$f$ and~$F$ satisfy~\eqref{fFdef} and~\eqref{fFinit} as well as the
regularity assumptions~\text{\rm{(r1)}} and~\text{\rm{(r2)}}.
Moreover, assume that the measures~$\tilde\rho_{s,t}$ given by~\eqref{rhost}
satisfy the weak EL equations~\eqref{ELtest_st} for all~$s$ and~$t$
(for jet spaces~$\tilde{\J}^\text{\rm{\tiny{test}}}_{s,t}$ chosen in agreement with~\eqref{JtestDef} and~\eqref{Cnontriv}).
Then for every compact~$\Omega \subset M$, the following expression vanishes:
\begin{align}
I_m^\Omega &:= \int_\Omega d\rho(x) \int_{M \setminus \Omega} d\rho(y)\,\big(\partial_{1,s} - \partial_{2,s} \big) \big( \partial_{1,t} + \partial_{2,t} \big)^{m-1} L\big(x_{s,t}, y_{s,t} \big) \Big|_{s=t=0}  \label{Ikdef0} \\
&\qquad \qquad \qquad \quad  -\frac{\nu}{2} \int_\Omega\partial_{s} \partial_{t}^{m-1}  f_{s,t}(x) \Big|_{s=t=0}\:  d\rho(x) 
= 0 \:. \label{Ikdef}
\end{align}
\end{Thm}
\noindent Here the indices of the symbols $\partial_{1,s}, \partial_{2,s}, \ldots$ again indicate that the derivatives
act only on the first or second argument of $L(x_{s,t},y_{s,t})$, respectively (note that, in
contrast to the derivative~$\nabla_{1,\tilde{\u}}$, the partial derivative~$\partial_{1,s}$ acts also on the
factor~$f_{s,t}$ in~\eqref{Ldef}).

Before giving the proof, we briefly explain the structure and significance of this result.
The double integral~\eqref{Ikdef0} is a surface layer integral~\eqref{IntrOSI}
and can thus be interpreted as a {\em{surface term}}.
The summand~\eqref{Ikdef}, on the other hand, is a {\em{volume term}}.
Therefore, the general structure of the above equation relating a surface term to a volume term
resembles the Gau{\ss} divergence theorem.
In many applications, the volume term vanishes.
Then, considering the situation that~$\Omega$ exhausts the region between two
surfaces which extend to spatial infinity (as explained in the introduction of~\cite{jet}; see~\cite[Figure~1]{jet}),
the surface layer integral does not depend on the choice of the surfaces,
thus giving rise to a {\em{conservation law}}.
If the volume term is non-zero, the above theorem still gives quantitative information on how the
surface layer integral depends on the choice of the surface.
In cases where the volume term has a physical interpretation, this gives insight
into the dynamics of the system (similar to the inhomogeneous Maxwell equations
in integral form). Moreover, an identity relating a surface to a volume term
is very useful for getting estimates similar to energy estimates for
hyperbolic PDEs~\cite{linhyp}.

\Proof[Proof of Theorem~\ref{thmosirho}] 
Using~\eqref{Cnontriv}, we conclude that the term in brackets in~\eqref{ELtest_st} vanishes, i.e.\
\[ \int_M L\big(x_{s,t}, y_{s,t} \big)\: d\rho(y) 
- \frac{\nu}{2}\: f_{s,t}(x) = 0 \]
for all~$s,t \in (-\delta, \delta)$ and all~$x \in M$. Since the right hand side is smooth in $t$ and $s$ by~\eqref{fFdef}, 
we can take $k:=m-1$ derivatives in $t$ and one derivative in $s$ to obtain
\beq\label{ProofIk4}
 \int_M  \big(\partial_{1,s}  + \partial_{2,s} \big) \big( \partial_{1,t} + \partial_{2,t} \big)^k  \: L\big(x_{s,t}, y_{s,t} \big)\: d\rho(y)  - \frac{\nu}{2}\: \partial_{s} \partial_{t}^k \: f_{s,t}(x) \Big|_{s=t=0}  = 0 \: ,
\eeq
where we exchanged differentiation and integration with the help of the regularity assumption~(r1). 
We next consider the weak EL equation~\eqref{ELtest_st}, evaluated at $s=0$ with the jet~$\tilde{\u} :=  \partial_s \big( f_{s,t}, F_{s,t} \big) \big|_{s=0}$. Assumption~(r2) assures that this is jet is in~$\tilde{\J}^\text{\rm{\tiny{test}}}_{0,t}$. Using that
\[ D_{1,u} \, \L\big(F_{0,t}(x) , F_{0,t}(y) \big) = \partial_{1,s} \, \L\big(F_{s,t}(x) , F_{s,t}(y) \big) \Big|_{s=0} \:, \]
the weak EL equation~\eqref{ELtest_st} can be written as
\[ \int_M \partial_{1,s}    \: L\big(x_{s,t}, y_{s,t} \big)\: d\rho(y)  - \frac{\nu}{2}\: \partial_{s} \: f_{s,t}(x) \Big|_{s=0}  = 0 \:, \]
valid for all $t$. We now differentiate~$k$ times with respect to~$t$, multiply by two and subtract~\eqref{ProofIk4}.
We thus obtain
\[ 
 \int_M  \big(\partial_{1,s}  - \partial_{2,s} \big) \big( \partial_{1,t} + \partial_{2,t} \big)^k  \: L\big(x_{s,t}, y_{s,t} \big)\: d\rho(y)  - \frac{\nu}{2}\: \partial_{s} \partial_{t}^k \: f_{s,t}(x) = 0 \:. \]
Integrating the last equation over~$\Omega$ gives
\beq \label{rel1}
\begin{split}
\int_\Omega & d\rho(x) \int_M d\rho(y) \: \big( \partial_{1,s} - \partial_{2,s} \big) \big( \partial_{1,v} + \partial_{2,v} \big)^k L\big(x_{s,t}, y_{s,t} \big) \\
&= \frac{\nu}{2} \int_\Omega \partial_{s} \partial_{v}^k  f_{s,t}(x)\: d\rho(x) \:.
\end{split}
\eeq
Since the integrand is anti-symmetric in its arguments~$x$ and~$y$, we also have
\[ 
\int_\Omega d\rho(x) \int_\Omega d\rho(y) \: \big( \partial_{1,s} - \partial_{2,s} \big) \big( \partial_{1,v} + \partial_{2,v} \big)^k L\big(x_{s,t}, y_{s,t} \big) = 0 \:. \]
Subtracting this equation from~\eqref{rel1} gives the result.
\QED

\begin{Example} {\bf{(Symmetry transformations)}} \label{exsymm} {\em{
We now illustrate the above theorem in the example of families of measures obtained
by applying symmetry transformations. Thus, specializing the setting
in~\cite[Section~3.1]{noether}, we let~$F$
be a smooth one-parameter family of diffeomorphisms
\[ 
F \in C^\infty \big( (-\delta, \delta) \times \F \rightarrow \F \big)
\quad \text{with} \quad F_0 = \1
\quad \text{and} \quad F_{-s} = F_s^{-1} \:. \]
We assume that~$F$ is a {\em{symmetry of the Lagrangian}}
in the sense that\footnote{This is a slightly stronger assumption than in~\cite{noether}, where we 
merely assume that~\eqref{symmlagr} holds for $x, y \in M$. We need the stronger assumption because
a necessary criterion for the weak EL equations~\eqref{ELtest_st} to hold is that term in brackets in~\eqref{ELtest_st} 
vanishes identically on $M$. In~\cite{noether}, however,
we merely used that the derivative of the term in brackets vanishes on $M$.}
\beq \label{symmlagr}
\L \big(x, F_s(y) \big) = \L \big(  F_{-s}(x), y \big)
\qquad \text{for all~$s \in (-\delta, \delta)$ and~$x, y \in \F$} \:.
\eeq
Moreover, we let~$\rho$ be a solution of the weak EL equations~\eqref{ELtest} and consider
the family of measures~$(\tilde \rho_{s,t})_{s,t \in (-\delta, \delta)}$ given by
\[ 
\tilde{\rho}_{s,t} := (F_s)_* \rho \]
(note that there is no $t$-dependence).
The symmetry assumption~\eqref{symmlagr} implies that the weak EL equations~\eqref{ELtest_st} are satisfied
for every measure $\tilde \rho_{s,t}$ (if the jet spaces~$\tilde{\J}^\text{\rm{\tiny{test}}}_{s,t}$
are identified by the symmetry transformation).
If the regularity assumptions~(r1) and~(r2) on page~\pageref{r1} hold, Theorem~\ref{thmosirho} applies.
Choosing~$m=1$, we obtain the conservation law
\begin{align*}  
0 &= \int_\Omega d\rho(x) \int_{M \setminus \Omega} d\rho(y) \;
\big(\partial_{1,s} - \partial_{2,s} \big) \,
\L\big( F_s(x), F_s(y) \big) \Big|_{s=0}  \\
&=  \frac{d}{ds}\int_\Omega d\rho(x) \int_{M \setminus \Omega} d\rho(y) \;
\Big( \L\big( F_s(x), y \big) - \L\big( F_{-s}(x), y \big) \Big)  \Big|_{s=0} \:.
\end{align*}
This equation agrees with the Noether-like theorem for a continuously differentiable symmetry
of the Lagrangian in~\cite{noether} (see~\cite[Theorem~3.3]{noether} and later generalizations in the same paper).
We conclude that the Noether-like theorem for a symmetry of the Lagrangian found in~\cite{noether}
is a corollary of the conservation laws found in Theorem~\ref{thmosirho}.
}} \QEDrem
\end{Example}

\subsection{Formulation in Terms of Linearized Solutions}\label{LinSol}
In order to work out applications of Theorem~\ref{thmosirho},
one must construct families of solutions of the weak EL equations~\eqref{ELtest_st}.
We now explain how this can be accomplished.
Our main result will be to rewrite the conserved surface layer integral of Theorem~\ref{thmosirho}
as an $m$-multilinear form on solutions of the linearized field equations (see Theorem~\ref{thmlin}).
Our results rely on the perturbation expansion for critical measures as developed in~\cite{perturb}.
In order to make the present paper self-contained, we recall the basics of the perturbation
expansion and work out the combinatorial details in Appendix~\ref{appcombi}.

In order to keep the setting as simple as possible, we always perform Taylor expansions
in components in given charts. Therefore, for any~$x \in M$ we choose a chart
of~$\F$ around~$x$  and work in components~$x^\alpha$. Also, we always write
the mapping~$F_{s,t}(x)$ in~\eqref{rhost} and vector fields on~$\F$ along~$M$
in this chart and expand componentwise. For ease in notation, we
usually omit the index~$\alpha$ from now on. But one should keep in mind that 
from now on we always work in suitably chosen charts.

Following~\cite{perturb}, we also introduce other jet spaces needed below. First, we define a 
convenient version of a dual of~$\Jtest$.
\begin{Def}\label{JteststarDef}
We denote the
continuous global one-jets of the cotangent bundle restricted to~$M$
by~$\J^* := C^0(M, \R) \oplus C^0(M, T^*\F)$.
We let~$(\Jtest)^*$ be the quotient space
\[ 
\begin{split}
(\Jtest)^* &:= \J^* \Big/  \big\{ (g,\varphi) \in \J^* \:\big|\: g(x) \,a(x) + \la \varphi(x), u(x) \ra = 0 \\
&\qquad\qquad\qquad\qquad\quad  \text{ for all~$\u=(a,u) \in \Jtest$ and~$x \in M$} \big\} \:,
\end{split} \]
where~$\la .,. \ra$ denotes the dual pairing of~$T^*_x\F$ and~$T_x\F$.
\end{Def} \noindent
Here, we have taken the quotient in order to avoid dual jets which are trivial on $\Jtest$.

We next specify the space of jets which can be used for varying the measure~$\rho$.
The basic idea is to define differentiability of the jets by corresponding
differentiability properties of the Lagrangian:
\begin{Def} \label{defJvary}
For any~$\ell \in \N_0  \cup \{\infty\}$, the jet space~$\J^\ell \subset \J$
is defined as the vector space of jets with the following properties:
\begin{itemize}[leftmargin=2em]
\item[\rm{(i)}] For all~$y \in M$ and all~$x$ in an open neighborhood of~$M$,
the directional derivatives
\beq \label{derex}
\big( \nabla_{1, \v_1} + \nabla_{2, \v_1} \big) \cdots \big( \nabla_{1, \v_p} + \nabla_{2, \v_p} \big) \L(x,y)
\eeq
(computed componentwise in charts around~$x$ and~$y$)
exist for all~$p \in \{1, \ldots, \ell\}$ and all~$\v_1, \ldots, \v_p \in \J^\ell$.
\item[\rm{(ii)}] The functions in~\eqref{derex} are $\rho$-integrable
in the variable~$y$, giving rise to locally bounded functions in~$x$. More precisely,
these functions are in the space
\[ L^\infty_\text{\rm{loc}}\Big( L^1\big(M, d\rho(y) \big), d\rho(x) \Big) \:. \]
\item[\rm{(iii)}] Integrating the expression~\eqref{derex} in~$y$ over~$M$
with respect to the measure~$\rho$,
the resulting function (defined for all~$x$ in an open neighborhood of~$M$)
is continuously differentiable in the direction of every jet~$\u \in \Jtest$.
\end{itemize}
\end{Def}

Working in charts also makes it possible to identify the tangent spaces at different
points simply by identifying the components. In particular, we use this
method in order to identify~$\tilde{\u}(F_{s,t}(x))$ with a jet~$\u(x)$.
We choose the jet space~$\tilde{\J}^\text{\rm{\tiny{test}}}_{s,t}$ such that
under this identification it coincides with~$\Jtest$. Then the family of
EL equations~\eqref{ELtest_st} simplifies to
\beq \label{final}
\nabla_{1,\u} \bigg( \int_M L\big(x_{s,t}, y_{s,t} \big) \, d\rho(y) - 
\frac{\nu}{2}\:f_{s,t}(x) \bigg) = 0 \qquad \text{for all~$\u \in \Jtest$} \:.
\eeq
We point out that by our conventions, the derivative~$\nabla_{1,\u}$ acts on the first argument of the Lagrangian~$\L$
and on the constant~$\nu$, but the factor~$f_{s,t}(x)$ is not differentiated.
We extend our conventions for partial derivatives and jet derivatives as follows:
\begin{itemize}[leftmargin=2em]
\itemD Partial and jet derivatives with an index $i \in \{ 1,2 \}$, as for example in~\eqref{derex}, only act on the respective variable of the function $\L$.
This implies, for example, that the derivatives commute,
\beq\label{ConventionPartial}
\nabla_{1,\v} \nabla_{1,\u} \L(x,y) = \nabla_{1,\u} \nabla_{1,\v} \L(x,y) \:.
\eeq
\itemD The partial or jet derivatives which do not carry an index act as partial derivatives
on the corresponding argument of the Lagrangian. This implies, for example, that
\[ \nabla_\u \int_\F \nabla_{1,\v} \, \L(x,y) \: d\rho(y) =  \int_\F \nabla_{1,\u} \nabla_{1,\v}\, \L(x,y) \: d\rho(y) \:. \]
\end{itemize}
We point out that, in contrast to the method and conventions used in~\cite{jet},
our conventions have the major benefit that {\em{jets are never differentiated}}.

In preparation, we need to define the notion of solutions of the linearized field equations
(for details see~\cite[Section~4.2]{jet}).
For~$\ell \in \N$ and $\v_1, \ldots , \v_\ell \in \J^\ell$, we define
\begin{align}\begin{split}\label{Boxl}
&\Delta_\ell \big[\v_1, \ldots, \v_\ell \big](x)
:= \frac{1}{\ell!} \bigg( \int_{M} \big( \nabla_{1, \v_1} + \nabla_{2, \v_1} \big) \cdots \big( \nabla_{1, \v_\ell} + \nabla_{2, \v_\ell} \big) \L(x,y)\: d\rho(y)  \\
&\qquad\qquad\qquad\qquad\qquad\quad -\frac{\nu}{2}\: b_1(x) \cdots b_\ell(x) \bigg) \:,
\end{split}\end{align}
where the $b_i$ denote the scalar components of $\v_i$, and where the space-time point~$x$ can be chosen
in a neighborhood of~$M \subset \F$.
According to the assumptions in Definition~\ref{defJvary}, the derivatives and the integral in~\eqref{Boxl} exist, and the
resulting function is continuously differentiable in the direction of any~$\u \in \Jtest$.
Therefore, the function~$\Delta_\ell \big[\v_1, \ldots, \v_\ell \big]$ can be identified
with a dual jet~$\w^* \in (\Jtest)^*$ (see Definition~\ref{JteststarDef}),
where the dual pairing with a jet~$\u \in \Jtest$ is given by
\beq \label{Pairing}
\la \u, \w^\ast \ra(x) := \nabla_\u \,\Delta_\ell \big[\v_1, \ldots, \v_\ell \big](x) \:.
\eeq
Using this identification, the operator~$\Delta_\ell$ in~\eqref{Boxl} gives rise to a mapping
\[ \Delta_\ell \::\: \underbrace{\J^\ell \times \cdots \times \J^\ell}_{\text{$\ell$ factors}}
\rightarrow (\Jtest)^* \:. 
\]
Using the abbreviation
\[ \Delta := \Delta_1 \: ,\]
the linearized field equations can be written as in the next definition (cf.~\cite[eq.~(4.19)]{jet}).

\begin{Def} \label{deflinear}
A jet~$\w \in \J^1$ is a {\bf{solution of the linearized field equations}}
if it satisfies the equation
\beq \label{lin}
\la \u, \Delta \w \ra = 0 \qquad \text{for all~$\u \in \Jtest$}\:.
\eeq
The vector space of all linearized solutions is denoted by~$\Jlin \subset \J^1$.
\end{Def} \noindent
For clarity, we point out that the equation in~\eqref{lin} is understood pointwise as
\[ \la \u, \Delta \w \ra(x) = 0 \qquad \text{for all~$x \in M$}\:. \]
Thus the linearized field equations are equations in space-time~$M$. 

The connection to families of solutions of the EL equations
is obtained as follows. Suppose that the family of measures~$\tilde{\rho}_{s,t}$
as defined by~\eqref{rhost} satisfies the EL equations~\eqref{final}
for all~$s$ and~$t$ (this entails that, as explained before~\eqref{final},
we identify the jet spaces~$\Jtest_{s,t}$ with~$\Jtest$ componentwise in our charts).
Then differentiating~\eqref{final} with respect to~$s$ or~$t$ at~$s=t=0$, one sees that
the corresponding jets given by
\[ \partial_s (f_{s,t}, F_{s,t})|_{s=t=0} \qquad \text{and} \qquad \partial_t (f_{s,t}, F_{s,t})|_{s=t=0} \]
are solutions of the linearized field equations~\eqref{lin}.
In other words, our family is described linearly in~$s$ and~$t$ by two solutions of the linearized
field equations. Likewise, the higher $s$-and $t$-derivatives of~\eqref{final} involve the
operator~$\Delta_\ell$ with~$\ell>1$. Setting these higher derivatives to zero gives rise to
equations which can be solved iteratively using the perturbation theory as developed systematically
in~\cite{perturb}. We now recall how this perturbation expansion works.
For clarity, we denote the expansion parameter by~$\lambda$.
In our situation, $\lambda$ can be thought of as describing the the size of~$s$ and~$t$.
This could be made precise by setting~$s = \lambda \hat{s}$ and~$t = \lambda \hat{t}$
with parameters~$\hat{s}$ and~$\hat{t}$ which are of the order one.
Then the limit~$s, t \rightarrow 0$ would correspond to taking the limit~$\lambda \rightarrow 0$.
For ease in notation, we avoid the hats and simply take~$\lambda$ as a formal parameter
which is used as a book-keeping device in order to keep track of the different orders in perturbation
theory. As we shall see, the conserved surface layer integral~$I^\Omega_m$ will involve
the perturbation expansion only up to the $m^\text{th}$ order. With this in mind,
it is unproblematic to work with formal power expansions in~$\lambda$, and
the convergence of the perturbation series is not an issue.
But one should keep in mind that our construction involves the assumption that there
exists a family~$(f_{s,t}, F_{s,t})$ which satisfies the regularity assumptions~(r1) and~(r2)
on page~\pageref{r1}.
These assumptions go beyond a perturbative treatment
(for the description on the level of formal power expansions see Section~\ref{secperturb} below).

Thus for the function~$f_{s,t}$ we make the power ansatz
\beq \label{fxser}
f_{s,t}(x) = \sum_{p=0}^\infty \lambda^p\: f_{s,t}^{(p)}(x)  \qquad \text{with} \qquad f_{s,t}^{(0)}(x) = 1 \:.
\eeq
For the expansion of~$F_{s,t}$, on the other hand, we choose a chart around~$x$ and
write~$F_{s,t}(x)$ in components as~$(F(x)^\alpha)_{\alpha=1,\ldots, m}$. Then we expand~$F_{s,t}$ componentwise,
\beq \label{Fxser}
F_{s,t}(x)^\alpha = \sum_{p=0}^\infty \lambda^p\: F_{s,t}^{(p)}(x)^\alpha
\qquad \text{with} \qquad F_{s,t}^{(0)}(x)^\alpha = x^\alpha \:.
\eeq
The choices of~$f_{s,t}^{(0)}$ and $ F_{s,t}^{(0)}$ ensure that for~$\lambda=0$,
the measure~$\tilde{\rho}$ in~\eqref{rhost} coincides with the unperturbed measure~$\rho$.
For ease in notation, we shall omit the index~$\alpha$ in the expansion of $F(x)$.
But one should keep in mind that the expansion of~$F(x)$ always involves the choice of a chart around~$x$.

It is preferable to write the function~$f_{s,t}$
in the family of measures~\eqref{rhost} as
\beq \label{cstdef}
f_{s,t} = \exp c_{s,t} \qquad \text{with} \qquad c \in C^\infty\big((-\delta, \delta)^2 \times \F,\R \big) \:.
\eeq
where the expansion of~$c_{s,t}$ in~$\lambda$ is denoted similar
to~\eqref{fxser} by
\[ c_{s,t}(x) = \sum_{p=0}^\infty \lambda^p\: c_{s,t}^{(p)}(x)  \qquad \text{with} \qquad c_{s,t}^{(0)}(x) = 0 \:. \]
We write the expansion coefficients again with jets,
\beq \label{vser}
\big( c_{s,t}, F_{s,t} \big)(x) = (0,x) + \sum_{p=1}^\infty \w_{s,t}^{(p)}(x) \:.
\eeq

In the next lemma we expand the formula in Theorem~\ref{thmosirho} in powers of~$\lambda$.
\begin{Lemma} \label{prpcombi} Let~$m \in \N$. If~\eqref{cstdef} and~\eqref{vser}
is the perturbation expansion of a family of measures~\eqref{rhost} which satisfies the weak EL
equations~\eqref{ELtest_st} and respects the regularity assumptions~{\rm{(r1)}} and~{\rm{(r2)}}, 
then for every compact $\Omega \subset M$ and every $p\geq0$, we have
\begin{align*}
0 &= I_{m,(p)}^\Omega := \sum_{\ell=1}^p \frac{1}{(\ell-1)!} \! \! \sum_{\stackrel{\text{\scriptsize{$q_1, \ldots, q_\ell \geq 1$}}} {\text{with } q_1+\cdots+q_\ell=p}} \ \
\sum_{\stackrel{\text{\scriptsize{$k_1, \ldots, k_\ell \geq 0$}}} {\text{with } k_1+\cdots+k_\ell=m-1}} \!\! \!\!\! {{m-1}\choose{k_1 \, \ldots \, k_\ell}} \\
&\qquad \times
\bigg( \int_\Omega d\rho(x) \int_{M \setminus \Omega} d\rho(y) \; 
\Big(\nabla_{1,\partial_s\partial_t^{k_1}\w_{s,t}^{(q_1)}} - \nabla_{2,\partial_s\partial_t^{k_1}\w_{s,t}^{(q_1)}} \Big) \\
&\qquad\qquad\qquad \times \Big(\nabla_{1,\partial_t^{k_2}\w_{s,t}^{(q_2)}} + \nabla_{2,\partial_t^{k_2}\w_{s,t}^{(q_2)}} \Big)
\cdots \Big(\nabla_{1,\partial_t^{k_\ell}\w_{s,t}^{(q_\ell)}} + \nabla_{2,\partial_t^{k_\ell}\w_{s,t}^{(q_\ell)}} \Big) \, \L(x,y)\\
&\qquad\qquad -\frac{\nu}{2} \int_\Omega \big(\partial_{s}  \partial_t^{k_1} c_{s,t}^{(q_1)}) \:\cdots\: (\partial_t^{k_\ell}  c_{s,t}^{(q_\ell)} \big) \:  d\rho(x) \bigg) \bigg|_{s=t=0}\:.
\end{align*}
\end{Lemma} \noindent
In order not to distract from the main ideas, the proof of this combinatorial lemma
is given in Appendix~\ref{appcombi}.

The higher order expansion terms in~\eqref{fxser} and~\eqref{Fxser}
can be expressed in terms of the linearized solutions using the Green's operator of the linearized field equations,
as we now explain (for the existence of Green's operators see~\cite{linhyp}).
\begin{Def} \label{defgreen}
A linear mapping~$S : (\Jtest)^* \rightarrow \J^\infty \cap \Jtest$
is referred to as a {\bf{Green's operator}} if
\beq \label{Sdefine}
\Delta\,S \, \v = -\v \qquad \text{for all~$\v \in (\Jtest)^*$} \:.
\eeq
\end{Def} \noindent
Similar as is the case for hyperbolic partial differential
equations or in relativistic physics, the Green's operators as defined here are not unique,
because adding a linearized solution to the Green's operator gives another Green's operator.
Consequently, there is a freedom in the choice of Green's operator to every order on perturbation theory.
We denote the corresponding choice by $S^{(p)}$. We thus obtain
\beq \label{vpiter}
\w_{s,t}^{(p)} = S^{(p)} \,E^{(p)} \:,
\eeq
where
\begin{align}
E^{(p)}(x) &= \sum_{\ell=2}^p E_\ell^{(p)}(x) \label{Ep} \\
E_\ell^{(p)}(x) &= \! \sum_{\stackrel{\text{\scriptsize{
$q_1, \ldots, q_\ell \geq 1$}}}{\text{with } q_1+\cdots+q_\ell=p}} \!
\Delta_\ell \big[ \w_{s,t}^{(q_1)}, \ldots, \w_{s,t}^{(q_\ell)} \big](x) \label{Elp} \:.
\end{align}
As shown in detail in~\cite{perturb}, this perturbation expansion gives rise to
families of solutions of the linearized field equations.

We now specify the dependence on the parameters~$s$ and~$t$. For~$\w^{(1)}_{s,t}$ we simply
take a linear combination of two solutions of the linearized field equations,
\beq \label{w1def} 
\w_{s,t}^{(1)} := s \,\u + t \,\v \qquad \text{with} \qquad \u, \v \in \Jlin \cap \Jtest \: .
\eeq
Computing the nonlinear corrections~$\w_{s,t}^{(p)}$ via~\eqref{vpiter}, we obtain a two-parameter
family of solutions of the weak EL equations. 
Setting~$t=0$, we denote the $p^\text{th}$ order correction to~$\v$
by~$\v^{(p)}:= \w_{s,t}^{(p)}|_{s=0}$.
The next lemma gives a connection between $t$-derivatives of $\w_{s,t}^{(q)}$
and the jets~$\v^{(k)}$.

\begin{Lemma}\label{tk}  For $q \geq 1$, we have
\[ 
\partial_t^k \w_{s,t}^{(q)} \big|_{s=t=0} = k! \, \v^{(k)} \delta_{q,k} \qquad\qquad \text{for $k\geq0$.} \]
\end{Lemma}
\Proof We proceed inductively in $q$. The statement holds for $q=1$ by~\eqref{w1def}. For the induction step $q-1 \rightarrow q$, 
we expand $\w^{(q)}$ according to~\eqref{vpiter},~\eqref{Ep} and~\eqref{Elp} as
\begin{align*}
\partial_t^k \:  \w_{s,t}^{(q)} \big|_{s=t=0} = S^{(q)}  \sum_{\ell=2}^q \sum_{\stackrel{\text{\scriptsize{
$q_1, \ldots, q_\ell \geq 1$}}}{\text{with } q_1+\cdots+q_\ell=q}}\!\!\!
\partial_t^k  \: \Delta_\ell \big[ \w_{s,t}^{(q_1)}, \ldots, \w_{s,t}^{(q_\ell)} \big](x)\big|_{s=t=0}
\end{align*}
where $\Delta_\ell$, using the notation
\[ 
\overline \nabla_{\w_{s,t}^{(q_\ell)} }  := \nabla_{1,  \w_{s,t}^{(q_\ell)} } + \nabla_{2,  \w_{s,t}^{(q_\ell)} } \:, \]
is given by
\begin{align}\label{w0wichtig}
\begin{split}
\Delta_\ell \big[ \w_{s,t}^{(q_1)},& \ldots, \w_{s,t}^{(q_\ell)}  \big](x)
 \\
&=  \frac{1}{\ell!} \bigg( \int_{M} \overline \nabla_{\w_{s,t}^{(q_1)}} \cdots \overline \nabla_{\w_{s,t}^{(q_\ell)} }\L(x,y)\: d\rho(y)  
 -\frac{\nu}{2}\: c_{s,t}^{(q_1)}(x) \cdots c_{s,t}^{(q_\ell)}(x) \bigg) \:.
\end{split}
\end{align}
Note that here, $q_i < q$ for $i=1,...,\ell$, making it possible to apply the induction hypothesis. This gives
\[ 
\partial_t^k  \:  \overline \nabla_{\w_{s,t}^{(q_1)}} \cdots \overline \nabla_{\w_{s,t}^{(q_\ell)} }\big|_{s=t=0}  \\
= k! \!\!\!\! \sum_{\stackrel{\text{\scriptsize{$k_1, \ldots, k_\ell \geq 0$}}} {\text{with } k_1+\cdots+k_\ell=k}} \!\!\! 
\overline \nabla_{\v^{(k_1)}} \cdots \overline \nabla_{\v^{(k_\ell)} } \ 
\delta_{k_1,q_1} \ldots \delta_{k_\ell,q_\ell} \]
and similarly in the second term in~\eqref{w0wichtig}. 
The inequalities~$q_1, ... , q_\ell \geq 1$ imply that in the last sum we get contributions
only if~$k_1, ... , k_\ell \geq 1$.
Combining the above equations proves the induction step.
This concludes the proof.
\QED

\begin{Lemma}\label{factors} The jet~$\w_{s,t}^{(q)}$ can be written as a sum, 
where each summand contains~$q$ factors $\w_{s,t}^{(1)}$.
\end{Lemma}
\Proof We again proceed by induction in~$q$. In the case~$q=1$, there is nothing to prove.
Thus assume that the statement holds for all~$q=1,\ldots, \tilde{q}-1$.
Then we can expand~$\w^{(\tilde{q})}$ according to~\eqref{vpiter},~\eqref{Ep} and~\eqref{Elp} to obtain
\begin{align*}
\w_{s,t}^{(\tilde{q})}
= S^{(\tilde{q})}  \sum_{\ell=2}^{\tilde{q}}  \sum_{\stackrel{\text{\scriptsize{
$q_1, \ldots, q_\ell \geq 1$}}}{\text{with } q_1+\cdots+q_\ell=\tilde{q}}}\!\!\! \Delta_\ell \big[ \w_{s,t}^{(q_1)}, \ldots, \w_{s,t}^{(q_\ell)} \big](x)
\end{align*}
where $\Delta_\ell$ is given by~\eqref{w0wichtig}. Using the induction hypothesis concludes the proof.
\QED

After these preparations, we can reformulate Theorem~\ref{thmosirho} in terms of
linearized solutions.
\begin{Thm} \label{thmlin}
Given~$m \in \N$, let $\w^{(1)}_{s,t}$ be as in~\eqref{w1def} and~$\w_{s,t}^{(q)}$
be the corresponding non-linear correction terms~\eqref{vpiter}--\eqref{Elp}.
If the resulting family satisfies the regularity assumptions~{\rm{(r1)}} and~{\rm{(r2)}}
on page~\pageref{r1}, then for every compact $\Omega \subset M$, we have
\begin{align}
0 & = I_m^\Omega := \sum_{\ell=0}^{m-1} \frac{(m-1)!}{ \ell! } \! \! \sum_{\stackrel{\text{\scriptsize{$q_1, \ldots, q_{ \ell+1} \geq 1$}}} {\text{with } q_1+\cdots+q_{\ell+1}=m}} 
\!\! \!\!\! \! \frac{1}{\: (q_1 - 1)!} \:
\notag \\
&\times\bigg(  \int_\Omega d\rho(x) \int_{M \setminus \Omega} d\rho(y)
\: \Big(\nabla_{1,\partial_s\partial_t^{(q_1 - 1)}\w_{s,t}^{(q_1)}|_{s=t=0} } - \nabla_{2,\partial_s\partial_t^{(q_1 - 1)}\w_{s,t}^{(q_1)}|_{s=t=0} } \Big) \: \notag \\[0.3em]
&\qquad\qquad\qquad\qquad\qquad \times \big(\nabla_{1,\v^{(q_2)}} + \nabla_{2,\v^{(q_2)}} \big) \: \cdots \big( \nabla_{1,\v^{(q_\ell)}} + \nabla_{2,\v^{(q_\ell)}} \big) \, \L(x,y) \notag \\[0.2em]
&\qquad -\frac{\nu}{2} \int_\Omega \big(\partial_{s}  \partial_t^{(q_1 - 1)} c_{s,t}^{(q_1)}|_{s=t=0} \big) \: b^{(q_2)}  \:\cdots\: b^{(q_\ell)}  \:  d\rho(x) \bigg) \:, \label{thmlinEq}
\end{align}
where we again denote the components of the jets by~$\v =(b,v)$.
\end{Thm}
\Proof 
Using Lemma~\ref{tk} and Lemma~\ref{prpcombi} gives
\begin{align*}
&I_{k+1,(p)}^\Omega = \sum_{\ell=1}^p \frac{k!}{(\ell-1)! } \! \! \sum_{\stackrel{\text{\scriptsize{$q_1, \ldots, q_\ell \geq 1$}}} {\text{with } q_1+\cdots+q_\ell=p}}  \ \
\sum_{\stackrel{\text{\scriptsize{$k_1\geq 0$}}} {\text{with } k_1+q_2+ \cdots+q_\ell=k}} \!\! \!\!\! \frac{1}{\: k_1!} \:
\Big( \int_\Omega d\rho(x) \int_{M \setminus \Omega} d\rho(y) \\
&\times (\nabla_{1,\partial_s\partial_t^{k_1}\w_{s,t}^{(q_1)}} - \nabla_{2,\partial_s\partial_t^{k_1}\w_{s,t}^{(q_1)}}) \: (\nabla_{1,\v^{(q_2)}} - \nabla_{2,\v^{(q_2)}}) 
\: \cdots \: (\nabla_{1,\v^{(q_\ell)}} + \nabla_{2,\v^{(q_\ell)}}) \, \L(x,y)\\
&\qquad -\frac{\nu}{2} \int_\Omega \big(\partial_{s}  \partial_t^{k_1} c_{s,t}^{(q_1)}) \: b^{(q_2)}  \:\cdots\: b^{(q_\ell)}  \:  d\rho(x) \Big) \Big|_{s=t=0} 
\end{align*}
The condition of the second summation implies that
\[ k_1 = k - (q_2 + \ldots + q_\ell) = k - p + q_1 \geq 0 \, .\]
Lemma~\ref{factors} implies that $\partial_s\partial_t^{k_1}\w_{s,t}^{(q_1)} \neq 0$ only if $q_1 = k_1 + 1$.
Thus, $I_{k+1,(p)}^\Omega \neq 0$ only if $m = k+1= p$.
Changing the summation parameter from $\ell$ to $\tilde \ell := \ell - 1$ gives the result.
\QED

\subsection{Corollaries and Discussion}\label{SecExamples}
We now discuss the implications of Theorems~\ref{thmosirho} and~\ref{thmlin}
in different situations. We will see that some of the resulting conservation laws
were discovered earlier in~\cite{noether} and~\cite{jet}.
But we will also find new conservation laws, such as
Theorem~\ref{thmintro} in the introduction. As before, we work with jets
denoted by~$\u = (a,u)$ and~$\v=(b,v)$.

We begin with the simplest case~$m=1$. In this case,
in~\eqref{thmlinEq} the dependence on $\v$ drops out and, with the help of~\eqref{w1def}, we obtain
\begin{align}
I_1^\Omega(\u) &= \int_\Omega d\rho(x) \int_{M \setminus \Omega} d\rho(y)\; 
\big( \nabla_{1, \partial_s \w_{s,t}^{(1)}} - \nabla_{2, \partial_s \w_{s,t}^{(1)}} \big) \, \L(x,y) \notag \\
& \qquad-\frac{\nu}{2} \int_\Omega \partial_s c_{s,t}^{(1)}(x) \bigg|_{s=t=0} d\rho(x) \notag \\
&= \int_\Omega d\rho(x) \int_{M \setminus \Omega} d\rho(y)\; 
\big( \nabla_{1, \u} - \nabla_{2, \u} \big) \, \L(x,y) -\frac{\nu}{2} \int_\Omega a(x)\:d\rho(x) \:. \label{I1osi}
\end{align}
The last integral is not a surface layer integral but a volume term.
For this reason, $I_1^\Omega$ in general does not give rise to a conservation law.
But it does if the scalar component of~$\u$ vanishes:

\begin{Corollary} {\bf{(The functional~$I^\Omega_1$)}} \label{ex0}
Assume that the jet~$\u$ is a solution of the linearized field equations
whose scalar component vanishes, i.e.
\[ \u=(0,u) \in \Jlin \cap \Jtest \:. \]
Then, setting~$m=1$, Theorem~\ref{thmosirho} gives rise to the
conserved surface layer integral
\beq \label{I0}
I_1^\Omega(\u) = \int_\Omega d\rho(x) \int_{M \setminus \Omega} d\rho(y)\; 
\big( D_{1, u} - D_{2, u} \big) \, \L(x,y)  \:.
\eeq
\end{Corollary} \noindent
This conservation law is reminiscent of the Noether-like theorems in~\cite{noether},
cf.\ Example~\ref{exsymm}. Indeed, if~$u$ is an infinitesimal generator of a symmetry of
the Lagrangian~\eqref{symmlagr}, the conservation of $I_1^\Omega(\u)$ 
reduces to these Noether-like theorems.
In particular, we thus recover current conservation
and the conservation of energy-momentum for Dirac wave functions
(see~\cite[Sections~5 and~6]{noether}).
However, the conservation law~\eqref{I0} goes beyond the Noether-like theorems
because it holds without any symmetry assumptions for any solution
of the linearized field equations whose scalar component vanishes.

Choosing~$m=2$ in~\eqref{thmlinEq}, 
and denoting $b^{(1)} = b$ and $\v^{(1)} = \v$ in accordance with~\eqref{w1def}, we obtain
\begin{align}
&I^\Omega_2(\u, \v) = \int_\Omega d\rho(x) \int_{M \setminus \Omega} d\rho(y)\;
\big( \nabla_{1, \partial_s \partial_t \w_{s,t}^{(2)}} - \nabla_{2, \partial_s \partial_t \w_{s,t}^{(2)}} \big) \:
\L(x,y) \Big|_{s=t=0} \notag \\
&\qquad -\frac{\nu}{2}\int_\Omega \partial_s \partial_t c_{s,t}^{(2)}(x) \Big|_{s=t=0} d\rho(x)  \notag \\
&\qquad + \int_\Omega d\rho(x) \int_{M \setminus \Omega} d\rho(y)\; 
\big( \nabla_{1, \partial_s \w_{s,t}^{(1)}} - \nabla_{2, \partial_s \w_{s,t}^{(1)}} \big) 
\big( \nabla_{1, \v} + \nabla_{2,\v} \big) \L(x,y) \Big|_{s=t=0} \notag \\
&\qquad -\frac{\nu}{2} \int_\Omega \big(\partial_s c_{s,t}^{(1)}(x)\big) \: b(x) \: d\rho(x) \Big|_{s=t=0} \notag \\
&= \int_\Omega d\rho(x) \int_{M \setminus \Omega} d\rho(y)\;
\big( \nabla_{1, \partial_s \partial_t \w_{s,t}^{(2)}} - \nabla_{2, \partial_s \partial_t \w_{s,t}^{(2)}} \big) \:
\L(x,y) \Big|_{s=t=0} \notag \\
&\qquad -\frac{\nu}{2}\int_\Omega \partial_s \partial_t c_{s,t}^{(2)}(x) \Big|_{s=t=0} d\rho(x) \notag \\
&\qquad + \int_\Omega d\rho(x) \int_{M \setminus \Omega} d\rho(y)\; 
\big( \nabla_{1, \u} - \nabla_{2, \u} \big) 
\big( \nabla_{1, \v} + \nabla_{2, \v} \big) \L(x,y) \notag \\
&\qquad -\frac{\nu}{2} \int_\Omega a(x) \: b(x)\: d\rho(x) \:, \label{I1omega}
\end{align}
where the term~$\partial_s \partial_t \w_{s,t}^{(2)}$ is given in second order perturbation theory by
\beq \label{I1Delta}
\partial_s \partial_t \w_{s,t}^{(2)} \big|_{s=t=0} = S^{(2)} \:
\partial_s \partial_t \Delta_2 \big[ \w_{s,t}^{(1)}, \w_{s,t}^{(1)} \big] \big|_{s=t=0}
= 2 \, S^{(2)} \, \Delta_2 \big[ \u, \v \big] \:,
\eeq
which is symmetric in $\u$ and $\v$ in view of~\eqref{ConventionPartial}.
Anti-symmetrizing in~$\u$ and~$\v$, the second order term
as well as the term involving~$\nu$ drop out. 

\begin{Corollary} {\bf{(The functional~$I^\Omega_2$ anti-symmetrized)}}\label{exsymp} 
Theorem~\ref{thmlin} gives rise to the conserved surface layer integral
\beq \label{I1omegaCons}
\sigma_\Omega(\u, \v) := \int_\Omega d\rho(x) \int_{M \setminus \Omega} d\rho(y)\;
\big( \nabla_{1, \u} \nabla_{2, \v} - \nabla_{1, \v} \nabla_{2, \u} \big) \, \L(x,y) \:.
\eeq
\end{Corollary}
\noindent This is precisely the symplectic form as found in~\cite{jet} (cf.~\cite[Section~3.3]{jet}).

We next symmetrize~$I_2^\Omega$ in its arguments~$\u$ and~$\v$.
Then~\eqref{I1omega} simplifies to
\begin{align}\begin{split}\label{I1symm}
&\frac{1}{2} \big(I_2^\Omega(\u, \v) + I_2^\Omega(\v, \u) \big) = \int_\Omega d\rho(x) \int_{M \setminus \Omega} d\rho(y) \,
\big( \nabla_{1, \u} \nabla_{1, \v} + 2 \,\nabla_{1, S^{(2)} \Delta_2[ \u, \v]} \\
& - \nabla_{2, \u} \nabla_{2, \v} - 2 \,\nabla_{2, S^{(2)} \Delta_2[ \u, \v]} \big) \L(x,y) 
-\int_\Omega \Big( a(x) \: b(x) + 2 \,\nabla_{S^{(2)} \Delta_2[\u,\v]} \Big) \frac{\nu}{2} \:d\rho(x) \: .
\end{split}\end{align}
This expression vanishes according to Theorem~\ref{thmlin}.
We thus obtain Theorem~\ref{thmintro} in the introduction, where we replaced the symbol
$S^{(2)}$ by $S$ and used that, according to our convention explained before~\eqref{ConventionPartial},
\[ \big( \nabla_{1,\u} \nabla_{1,\v} \, \frac{\nu}{2} \big)(x) = a(x) \,  b(x) \, \frac{\nu}{2} \, .\]

As in the above example for~$m=1$, the last integral in~\eqref{I1symm} is a volume term.
In order to obtain a conservation law, we need to assume that the scalar component
of at least one jet vanishes, i.e.
\beq \label{ScalarVanishAsmpt}
\u = (0, u) \qquad \text{or} \qquad \v = (0,v)\:. 
\eeq
Moreover, we need to assume that the Green's operator gives no scalar component, i.e.
\beq\label{AssumptionGreenScalar}
S^{(2)} \Delta_2[ \u, \v] = (0, w) \qquad \text{with} \quad w \in \Gamma \:. 
\eeq
Under these additional assumptions, we obtain:
\begin{Corollary} {\bf{(The functional~$I^\Omega_2$ symmetrized)}} \label{exsprod} 
If~\eqref{ScalarVanishAsmpt} and~\eqref{AssumptionGreenScalar} hold,
Theorem~\ref{thmlin} gives rise to the conserved surface layer integral
\[ \begin{split} 
&\int_\Omega d\rho(x) \int_{M \setminus \Omega} d\rho(y)\;
\big( D_{1, u} D_{1, v} + 2 \,\nabla_{1, S \Delta_2[ \u, \v]} \\
& \qquad \qquad \qquad \qquad \qquad
- D_{2, u} D_{2, v} - 2 \,\nabla_{2, S \Delta_2[ \u, \v]}
\big) \, \L(x,y) \:.
\end{split} \]
\end{Corollary}
It is worth noting that, in contrast to the symplectic form of Corollary~\ref{exsymp},
the conserved surface layer integral involves a nonlinear correction term.

\subsection{Dependence on the Green's Operators}\label{DepGreensOp} $\;\;$
According to Theorem~\ref{thmlin} and~\eqref{vpiter}, the
functionals~$I^\Omega_m$ depend on the choice of the Green's
operators~$S^{(p)}$.  At first sight, this freedom might be surprising because conserved physical quantities like
currents or energy-momentum are canonically defined and do not involve any
arbitrariness. In this section we will analyze what this arbitrariness means.
In short, the answer is that different choices of the Green's operators correspond to taking different
linear combinations of conserved quantities. Before working out this statement systematically, we give a simple example.

\begin{Example} \label{ex1green}
{\em{ We return to the functional~$I^\Omega_2$
as considered in the above Corollaries~\ref{exsymp} and~\ref{exsprod}.
Clearly, this functional depends on the choice of the Green's operator~$S^{(2)}$, because
the jet~$\w_{s,t}^{(2)}$ in~\eqref{I1omega} is given by
\[ \w_{s,t}^{(2)} = \big( c_{s,t}^{(2)}, w_{s,t}^{(2)} \big) = S^{(2)} \Delta_2 \big(\w_{s,t}^{(1)}, \w_{s,t}^{(1)} \big) \:. \]
As explained after Definition~\ref{defgreen}, Green's operators are unique only up to solutions of the linearized
field equations. Therefore, we can modify~$S^{(2)}$ according to
\beq \label{S2trans}
S^{(2)} \rightarrow S^{(2)} + K^{(2)}\:,
\eeq
where~$K^{(2)}  : (\Jtest)^* \rightarrow \Jlin \cap \Jtest$ maps to the linearized solutions.
Then, according to~\eqref{I1omega} and~\eqref{I1Delta},
\begin{align*}
&I^\Omega_2(\u,\v) \rightarrow \; I^\Omega_2(\u,\v) - 2 \int_\Omega \nabla_{K^{(2)} \Delta_2[\u,\v]} \: \frac{\nu}{2}\: d\rho(x) \\
&\qquad + 2 \int_\Omega d\rho(x) \int_{M \setminus \Omega} d\rho(y)\;
\big( \nabla_{1, K^{(2)} \Delta_2[\u,\v]} - \nabla_{2, K^{(2)} \Delta_2[\u,\v]} \big) 
\L(x,y)  \\
&= I^\Omega_2(\u,\v) + 2 \, I^\Omega_1 \big(K^{(2)} \Delta_2[\u,\v] \big) \:.
\end{align*}
Therefore, the functional~$I^\Omega_2$ is modified by the functional~$I^\Omega_1$,
evaluated for the linearized solution~$K^{(2)} \Delta_2[\u,\v]$.
}} \QEDrem
\end{Example}

We next analyze the dependence on the Green's operators systematically.
We denote the conserved surface layer integrals~$I^\Omega_m$ expressed in terms
of linearized solutions~$\u, \v \in \Jlin \cap \Jtest$ by~$I^\Omega_m(\u,\v)$
(see Theorem~\ref{thmlin} and~\eqref{w1def}).
Clearly, these conserved surface layer integrals depend on the choice of the Green's operators.
In generalization of~\eqref{S2trans} we now modify the Green's operator of order~$q \in \{2,3,\ldots\}$ by
\beq \label{Schange1}
S^{(q)} \rightarrow \hat{S}^{(q)} := S^{(q)} + K^{(q)} \qquad \text{with} \qquad
K^{(q)} \::\: (\Jtest)^* \rightarrow \Jlin \cap \Jtest \:,
\eeq
leaving all the other Green's operators unchanged,
\beq \label{Schange2}
\hat{S}^{(\ell)} = S^{(\ell)} \qquad \text{for all~$\ell \neq q$} \:.
\eeq
We denote the conserved surface layer integrals corresponding to the modified Green's operators by~$\hat{I}^\Omega_m(\u,\v)$.

\begin{Thm} \label{thmgreen}
Modifying the Green's operators according to~\eqref{Schange1} and~\eqref{Schange2},
the conserved surface layer integrals for~$k \in \N_0$ transform according to
\beq \label{Ikgreen}
\begin{split}
\hat{I}^\Omega_{k+1}(\u,\v)
= \sum_{\ell=0}^k \begin{pmatrix} k \\ \ell \end{pmatrix} \bigg( \frac{d}{dt} \bigg)^{k-\ell}
I_{\ell+1}^\Omega \Big(\u + &q \,t^{q-1}\, K^{(q)} E^{(q)}(\u, \v, \ldots, \v),\\
&\: 
\v + t^{q-1}\, K^{(q)} E^{(q)}(\v,\v, \ldots, \v) \Big) \Big|_{t=0} \:,
\end{split}
\eeq
where we consider the inhomogeneity of the $q^\text{th}$ order in perturbation theory (see~\eqref{Ep})
as a symmetric $q$-linear functional on the linearized solutions,
\[ E^{(q)} \::\:  \underbrace{(\Jlin \cap \Jtest) \times \cdots \times (\Jlin \cap \Jtest)}_{\text{$q$ factors}}
\rightarrow (\Jtest)^* \:. \]
\end{Thm}
\Proof It is convenient to combine all the conserved surface layer integrals of Theorem~\ref{thmosirho}
in a {\em{generating functional}}~$I^\Omega_t$, being defined as the formal power series
\begin{align*}
I_t^\Omega &= \sum_{k=0}^\infty \frac{t^k}{k!}\: I^\Omega_{k+1} \\
&=\int_\Omega d\rho(x) \int_{M \setminus \Omega} d\rho(y)\,\big(\partial_{1,s} - \partial_{2,s} \big) 
L\big(x_{s,t}, y_{s,t} \big) \Big|_{s=0} 
-\frac{\nu}{2} \int_\Omega\partial_{s} f_{s,t}(x) \Big|_{s=0}\:  d\rho(x) \:.
\end{align*}
This surface layer integral depends on a two-parameter family of critical measures~$\tilde{\rho}_{s,t}$.
In Section~\ref{LinSol} we expressed this family in terms of linearized solutions~$\u$ and~$\v$
(see Theorem~\ref{thmlin} and~\eqref{w1def}). We denote the above generating functional
expressed in terms of the linearized solutions by
\beq \label{Iex}
I^\Omega(\u, t \v) := \sum_{k=0}^\infty \frac{t^k}{k!}\: I^\Omega_{k+1}(\u,\v) \:.
\eeq
The individual surface layer integrals are recovered by differentiating,
\[ I^\Omega_{k+1}(\u,\v) = \frac{d^k}{dt^k} I^\Omega(\u, t \v) \Big|_{t=0} \:. \]
Moreover, we denote the formal perturbation expansion for linearized solutions compactly by~$\Pert$, i.e.
\[  \Pert \::\: \Jlin \cap \Jtest  \rightarrow \J^\infty \cap \Jtest \:,\qquad
\w^{(1)} \mapsto \sum_{p=1}^\infty \w^{(p)}(x) \:. \]
Then the family of solutions~$\w_{s,t}$ constructed in Section~\ref{LinSol} can be written as
\[ \w_{s,t} = \Pert(s \,\u + t \,\v)\:. \]
The corresponding objects defined for the modified Green's operators
are denoted by an additional hat.

We now consider a family of critical measures~$\rho_{s,t}$ corresponding to the modified
perturbation expansion~$\hat{\Pert}(s\,\u + t \v)$. Assume that this family of nonlinear solutions can also
be expressed in terms of the unmodified Green's operators by
\[ \hat{\Pert}(s\,\u + t \v) = \Pert\big(s\,\u_{s,t} + t \,\v_{s,t} \big) \:, \]
where~$\u_{s,t}$ and~$\v_{s,t}$ are families of linearized solutions which depend on the parameters~$s$ and~$t$,
\[ \u_{s,t}, \v_{s,t} \in \Jlin \cap \Jtest \]
(these families will be constructed in detail below). Then
\beq \label{hI}
\hat{I}^\Omega_{k+1}(\u,\v) = \frac{d^k}{dt^k} \hat{I}^\Omega(\u, t \v) \Big|_{t=0}
= \frac{d^k}{dt^k} I^\Omega_t \Big|_{t=0} \:.
\eeq
On the other hand, expressing~$I^\Omega_t$ in terms of the unmodified perturbation expansion,
we can fix the parameters~$s$ and~$t$ of the subscripts
and expand only in the prefactor~$t$,
\[ I_t^\Omega = I^\Omega\big(\u_{0,t}, t \,\v_{0,t} \big)
\overset{\eqref{Iex}}{=} \sum_{\ell=0}^\infty \frac{t^\ell}{\ell!} \:I_{\ell+1}^\Omega\big(\u_{0,t}, \v_{0,t} \big) \:. \]
Using this relation in~\eqref{hI} gives
\begin{align}
\hat{I}^\Omega_{k+1}(\u,\v) &= 
\frac{d^k}{dt^k} \sum_{\ell=0}^\infty \frac{t^\ell}{\ell!} \:I_{\ell+1}^\Omega\big(\u_{0,t}, \v_{0,t} \big) \Big|_{t=0} \notag \\
&= \sum_{\ell=0}^k \begin{pmatrix} k \\ \ell \end{pmatrix} \bigg( \frac{d}{dt} \bigg)^{k-\ell}
I_{\ell+1}^\Omega\big(\u_{0,t}, \v_{0,t} \big) \Big|_{t=0} \:. \label{Iexp}
\end{align}

It remains to compute the families of linearized solutions~$\u_{s,t}$ and~$\v_{s,t}$.
We expand in powers of~$s \u+t \v$.
Clearly, up to the order~$q-1$ in perturbation theory, the perturbation expansions with and without hat
coincide. To the order~$q$, the modification of the Green's operator
can be compensated by changing the linearized solutions.
Indeed, according to~\eqref{Schange1} and~\eqref{vpiter}, by choosing
\beq \label{uvchoice}
s \,\u_{s,t} + t \,\v_{s,t} = s \,\u + t \,\v + K^{(q)} E^{(q)}\big( \underbrace{s\u+t \v, \ldots, s\u+t\v}_{\text{$q$ factors}} \big)
\eeq
we can arrange that the perturbation expansions coincide up to the order~$q$.
Since to every order higher than~$q$, the Green's operators again coincide~\eqref{Schange2},
the perturbation expansions with and without hat coincide to all orders.
We conclude that the families of linearized solutions~$\u_{s,t}$ and~$\v_{s,t}$
must satisfy~\eqref{uvchoice} as a formal power series in~$s$ and~$t$. This leads us to
\[ \u_{0,t} = \u + q\, K^{(q)} E^{(q)}(\u, t \v,\ldots, t \v) \qquad \text{and} \qquad
\v_{0,t} = \v + t^{q-1}\, K^{(q)} E^{(q)}(\v, \ldots, \v) \]
(where for convenience we chose~$\partial_s \v_{s,t}|_{s=0}=0$).
Using these formulas in~\eqref{Iexp} gives the result.
\QED

The proof of this theorem also explains how
the relation between different conservation laws comes about:
Different perturbation expansions give different ways of describing
nonlinear solutions in terms of linearized solutions.
Visualizing the nonlinear solution space as a (possibly infinite-dimensional) manifold,
different perturbation expansions can be regarded as different local charts.
The derivatives with respect to the parameters~$s$ and~$t$ correspond to
directional derivatives computed in the charts.
Higher directional derivatives are no tensors, meaning that when transforming
them from one chart to another, derivatives of the transformation mappings
come into play. This is how the transformation law~\eqref{Ikgreen} can be understood from a differential
geometric perspective.

\subsection{Perturbative Description} \label{secperturb}
In the applications, the perturbation expansion is usually treated as a formal power expansion in
a real parameter~$\lambda$, which can be interpreted as the coupling constant
(for details see~\cite{perturb} and the application to the bosonic Fock space
description in~\cite{fockbosonic}).
We now specify the regularity assumptions for the jets which seem sufficiently general for most
applications and ensure that the conditions~(r1) and~(r2) 
on page~\pageref{r1} are satisfied to every order in perturbation theory.

\begin{Def} \label{defslr}
The jet space~$\Jtest$ is {\bf{surface layer regular}} of order~$m \in \N$
if~$\Jtest \subset \J^m$ and
if for all~$\u, \v \in \Jtest$ and all~$p \in \{1, \ldots, m\}$ the following conditions hold:
\begin{itemize}[leftmargin=2em]
\item[\rm{(i)}] The directional derivatives
\beq \nabla_{1,\u} \,\big( \nabla_{1,\v} + \nabla_{2,\v} \big)^{p-1} \L(x,y) \label{Lderiv1}
\eeq
exist.
\item[\rm{(ii)}] The functions in~\eqref{Lderiv1} are $\rho$-integrable
in the variable~$y$, giving rise to locally bounded functions in~$x$. More precisely,
these functions are in the space
\[ L^\infty_\text{\rm{loc}}\Big( L^1\big(M, d\rho(y) \big), d\rho(x) \Big) \:. \]
\item[\rm{(iii)}] The $\u$-derivative in~\eqref{Lderiv1} may be interchanged with the $y$-integration, i.e.
\[ \int_M \nabla_{1,\u} \,\big( \nabla_{1,\v} + \nabla_{2,\v} \big)^{p-1} \L(x,y)\: d\rho(y)
= \nabla_{1,\u} \int_M \big( \nabla_{1,\v} + \nabla_{2,\v} \big)^{p-1} \L(x,y)\: d\rho(y) \:. \]
\end{itemize}
\end{Def} \noindent
Note that this is similar to Definition~\ref{defJvary}. The main structural difference is that
we now demand that the $\u$-derivative in~\eqref{Lderiv1} exists for all~$x$ and~$y$, whereas in
Definition~\ref{defJvary}~(iii) were merely demand that the $y$-integral is differentiable.

\begin{Prp} Assume that~$\Jtest$ is surface layer regular. Then, treating the perturbation expansion
as a formal power expansion and choosing a fixed jet space for testing (i.e.\ $\Jtest_{s,t} = \Jtest$
for all~$s$ and~$t$), then the regularity conditions~{\rm{(r1)}} and~{\rm{(r2)}} 
on page~\pageref{r1} are satisfied to every order in perturbation theory.
\end{Prp}
\Proof According to our assumptions on the Green's operators, the jets~$\w^{(p)}$ are all
in~$\Jtest$ (cf.~\eqref{vpiter} and Definition~\ref{defgreen}).
Therefore, the condition~\eqref{r2Eq} holds to every order in perturbation theory, proving
that~(r2) holds. In order to prove~(r1), we first note that the $s$- and $t$-derivatives
in~\eqref{r1Eq} (which act on both arguments of~$\L$) can be obtained as the contributions
of order~$m-1+q$ to a perturbation expansion. Therefore, restricting attention to
this order in perturbation theory, we obtain automatically that the integrals and the derivative
commute, i.e.\
\[ \partial_s^q \partial_t^{m-1}  \int_M L\big( x_{s,t}, y_{s,t} \big) \:d\rho(y) \Big|_{s=t=0}
= \int_M \partial_s^q \partial_t^{m-1} \, L\big( x_{s,t}, y_{s,t} \big) \Big|_{s=t=0} \:d\rho(y) \:. \]
Computing the integrand perturbatively, we obtain a sum of terms of the form
as in~\eqref{derex}. Therefore, the integral is well-defined according to 
the assumptions~(i) and~(ii) in Definition~\ref{defJvary}.
Moreover, in the case~$p=1$, we may use the assumptions~(i) and~(iii)
in Definition~\ref{defslr} to take the $\u$-derivative and to interchange it with
the integral. This concludes the proof.
\QED

\section{Example: A Lattice System} \label{seclattice}
We now illustrate the previous constructions in a detailed example 
introduced in~\cite[Section~5]{jet}. In this example, the minimizing measure
is supported on a two-dimensional lattice in Minkowski space.
The dynamical degrees of freedom are an~$S^1$-valued field on the lattice.
The Euler-Lagrange equations for this field give the discrete wave equation on the lattice.

After recalling the definition of the model, we construct the Green's operators.
Computing the surface layer integrals of Corollaries~\ref{ex0} and Theorem~\ref{thmintro},
we find that~$I_1$ vanishes identically, whereas~$I_2$ gives rise to a
non-trivial conserved quantity (the symplectic form in Corollary~\ref{exsymp} was already
computed in~\cite[Section~5.4]{jet}).

\subsection{Definition}
Let~$(\R^{1,1}, \la .,. \ra)$ be two-dimensional Minkowski space. Denoting two
space-time points by~$\underline{x} = (x^0, x^1)$ and~$\underline{y} = (y^0, y^1)$, the inner product
takes the form
\[ \la \underline{x}, \underline{y} \ra = x^0 y^0 - x^1 y^1 \:. \]
Let~$\F$ be the set
\[ \F = \R^{1,1} \times S^1\:. \]
We denote points in~$x \in \F$ by $x=(\underline{x}, x^\varphi)$
with~$\underline{x} \in \R^{1,1}$ and~$x^\varphi \in [-\pi, \pi)$.
Next, we let~$A$ be the square
\[ A = (-1,1)^2 \subset \R^{1,1} \:. \]
Moreover, given~$\varepsilon \in (0,\frac{1}{4})$,
we let~$I$ be the the following subset of the interior of the light cones,
\[ I = \big\{ \underline{x} \in \R^{1,1} \, \big| \, \la \underline{x}, \underline{x} \ra > 0
\textrm{ and } |x^0| < 1 +\varepsilon \big\} \:. \]
Let~$f : \R^{1,1} \rightarrow \R$ be the function
\[ f(\underline{x}) = \chi_{B_\varepsilon(0,1)}(\underline{x}) + \chi_{B_\varepsilon(0,-1)}(\underline{x})
- \chi_{B_\varepsilon(1,0)}(\underline{x}) - \chi_{B_\varepsilon(-1,0)}(\underline{x}) \:, \]
where~$\chi$ is the characteristic function and~$B_\varepsilon(x^0,x^1)$ denotes the open Euclidean ball 
of radius~$\varepsilon$ centred around $(x^0,x^1) \in \R^2 \simeq \R^{1,1}$. Finally, we let~$V : S^1 \rightarrow \R$ be the function
\[ V(\varphi) = 1 - \cos \varphi \:. \]
Given parameters~$\delta > 0$, $\lambda_I \geq 2$ and
\beq \label{lamin}
\lambda_A \geq 2 \lambda_I+\varepsilon \:,
\eeq
the Lagrangian~$\L$ is defined by
\beq
\begin{split}
\L(x,y) &= \lambda_A \ \chi_{A}( \underline x - \underline y) + \lambda_I \ \chi_{I}( \underline x - \underline y) +
V(x^\varphi -y^\varphi) \: f(\underline x - \underline y) \\
&\quad + \delta \: \chi_{B_\varepsilon(0,0)}(\underline{x}-\underline{y})\: V(x^\varphi -y^\varphi )^2 \:.
\end{split} \label{DefL}
\eeq
The following results have been proven in~\cite[Section~5]{jet}.

\begin{Lemma} \cite[Lemma~5.1]{jet}
The function~$\L(x,y)$ is non-negative and satisfies the conditions~\rm{(i)} and \rm{(ii)} on page~\pageref{Cond1}.
\end{Lemma}

\noindent We  introduce a measure~$\rho$ supported on the unit
lattice~$\UL:= \Z^2 \subset \R^{1,1} \subset \F$ by
\beq
\label{DefRho}
\rho =  \sum_{\underline x \in \UL} \: \delta_{(\underline x,0)} \:,
\eeq
where $\delta_{(\underline x, x^\varphi)}$ denotes the Dirac measure at $(\underline x, x^\varphi) \in \F$.

\begin{Lemma} \cite[Lemma~5.2]{jet}
The measure~$\rho$ satisfies the condition~\rm{(iii)} on page~\pageref{Cond4}.
\end{Lemma}

\noindent Clearly, the support of $\rho$ is given by
\beq \label{Mlattice}
M:= \supp \rho = \UL \times \{0\} \subset \F \:.
\eeq

\begin{Lemma}\label{ExStrongEL} \cite[Lemma~5.3]{jet}
The measure~\eqref{DefRho}
satisfies the EL equations~\eqref{ELstrong} if the parameter~$\nu$ in~\eqref{elldef} is chosen as
\beq \label{nuval}
\nu = 2 \lambda_A + 4 \lambda_I \:.
\eeq
\end{Lemma}

We remark that the measure $\rho$ is a local minimizer of the causal action as defined in~\cite[Proposition~4.10]{jet};
see~\cite[Corollary~5.5]{jet}.

\subsection{The Jet Spaces}
We next determine the jet spaces. Recall that in~\eqref{SmoothVecFields}, the smooth vector fields on $M$ 
were defined as those
vector fields which can be extended smoothly to $\F$. In our setting of a discrete lattice~\eqref{Mlattice}, this is
the case for every vector field.
Thus the jet space~\eqref{Jdef} can be written as
\[ \J = \big\{ \u = (a,u) \text{ with } a : M \rightarrow \R \text{ and } u : M \rightarrow T\F \big\} \:. \]
We denote the vector component by $u=(u^0, u^1, u^\varphi)$.
In order to determine the differentiable jets, we recall from the the proof of~\cite[Lemma~5.3]{jet} that
\[ \left\{ \begin{array}{ll}
\ell(\underline x, x^\varphi) =  \delta \, V(x^\varphi)^2 & \textrm{if }\underline{x} \in \UL \\[0.3em]
\ell(\underline x, x^\varphi) \geq \lambda_A + \delta \, V(x^\varphi)^2 &  \textrm{if } \underline{x} \notin \UL \:,
\end{array} \right. \]
where $\ell$ is defined as in~\eqref{elldef}. Hence the differentiable jets~\eqref{Jdiff} are given by
\beq\label{ExJDiff}
\Jdiff = \big\{ \u = (a,u) \text{ with } a : M \rightarrow \R \text{ and } u=(0,0,u^\varphi) : M \rightarrow T\F \big\} \:.
\eeq
We choose
\[ 
\Jtest = \Jdiff \:, \]
which implies that
\[ (\Jtest)^\ast =  \big\{ \v = (b,v) \text{ with } b : M \rightarrow \R \text{ and } v=(0,0,v^\varphi) : M \rightarrow T^\ast\F \big\} \:. \]
The proof of~\cite[Proposition~5.6]{jet} shows that $\v=(b,v) \in \J^1$ as in Definition~\ref{defJvary}
consist of an arbitrary scalar component $b$ and the vector component
\[ v(x) = \big(\underline{v}, v^\varphi(\underline x) \big) \:, \]
where $\underline{v}$ is a constant vector in $\R^{1,1}$ and $v^\varphi : M \rightarrow \R$ is an arbitrary function.
Since the derivative in direction of the constant vector vanishes, the resulting term reads
\[ 
\big( \nabla_{1, \v} + \nabla_{2, \v} \big) \, \L(x,y) = \big( b(\underline{x}) + b(\underline{y}) + v^\varphi(\underline x) \, \partial_{x^\varphi} + v^\varphi(\underline y) \, \partial_{y^\varphi} \! \big) \, \L(x,y) \:. \]
Our convention~\eqref{ConventionPartial} implies that the jet derivatives in~\eqref{derex} act on $\L(x,y)$ but not on 
the other jets. Since $\L(x,y)$ is smooth
in $x^\varphi$ and $y^\varphi$, it follows that the conditions for $\v$ being an element of $\J^\ell$ for $\ell >1$ 
are not stronger than for the case $\ell=1$.
Hence we conclude that
\[ \J^l = \J^1 \qquad \text{ for all } l \in \N_0  \cup \{\infty\} \: . \]

Finally, the solutions of the linearized field equations~\eqref{lin} are characterized in the following proposition.
We use the notation  $e_t := (1,0) \in \R^{1,1}$.
\begin{Prp}\label{ExLin} \cite[Proposition~5.6]{jet}
The linearized solutions $\Jlin$ of Definition~\ref{deflinear} consist of all
jets $\v = (b,v) \in \J$ with the following properties:
\begin{itemize}
\item[{\rm{(A)}}] The scalar component~$b : M \rightarrow \R$ satisfies the equation
\[ 
\lambda_A \: b(\underline x,0) + \lambda_I\, \big( b(\underline x + e_t,0) + b(\underline x - e_t,0)
\big) = 0 \:. \]
\item[{\rm{(B)}}] The vector component~$v : M \rightarrow T\F$ consists of a constant
vector~$\underline{v} \in \R^{1,1}$ and a function~$v^\varphi : M \rightarrow \R$, i.e.
\beq\label{ExVecComp}
v(x) = \big(\underline{v}, v^\varphi(x) \big) \:,
\eeq
where the function~$v^\varphi$ satisfies the discrete wave equation on $\UL$,
\beq \label{ExDiscrWaveEq}
\sum_{\underline y \in \UL}  f(\underline x - \underline y) \, v^\varphi\!(\underline y, 0) =  0 \, .
\eeq
\end{itemize}
\end{Prp}

\subsection{The Green's Operators}\label{ExGreens}
In this section we determine the Green's operators according to Definition~\ref{defgreen}
and analyze the regularity assumptions~(r1) and~(r2) on page~\pageref{r1}.
For~$\u=(a,u)\in\Jtest$ and $\v=(b,v) \in  \J^\infty$, according to~\eqref{Boxl} and~\eqref{Pairing}
the linearized field equations read
\begin{align}
\label{ExWeakEL}
\begin{split}
\la \u, \Delta \v \ra |_x&=  \big( \lambda_ A + 2 \lambda_I \big) \,a(\underline{x})\, b(\underline{x}) +
\lambda_A \,a(\underline{x}) \,b(\underline{x}) + \lambda_I \big( a(\underline{x}) \,b(\underline{x}+e_t)\\
&\qquad + a(\underline{x}) \,b(\underline{x}-e_t) \big) 
- u^\varphi\!( \underline x) \sum_{\underline y \in \UL} v^\varphi\!(\underline y)  \, f(\underline x - \underline y)     - a(\underline x) \,b(\underline x) \:\frac{\nu}{2} \:.
\end{split}
\end{align}
Hence, denoting the scalar and vector components of $S \v$ by $(sb,sv)$ and using~\eqref{nuval}, 
the defining equation for the Green's operator~\eqref{Sdefine} reads
\begin{align}
\label{ExGreensCond}
\begin{split}
& \lambda_A \,a(\underline{x}) \,sb(\underline{x}) + \lambda_I \big( a(\underline{x}) \,sb(\underline{x}+e_t) + a(\underline{x}) \,sb(\underline{x}-e_t) \big) \\
& -  u^\varphi\!( \underline x) \sum_{\underline y \in \UL} sv^\varphi\!(\underline y)  \, f(\underline x - \underline y) 
\stackrel{!}{=} - \la \u, \v \ra |_x =  - a(\underline x) \, b(\underline x)  - u^\varphi(\underline x) v^\varphi(\underline x)
\end{split}
\end{align}
for all $\u \in \Jtest$ and $\v \in (\Jtest)^\ast$.
Choosing~$\u = (a,0) \in \Jtest$ with arbitrary $a$, we have
\begin{align}
\label{ExGreensCond2}
\begin{split}
& \lambda_ A \, sb(\underline{x}) + \lambda_I \big( \,sb(\underline{x}+e_t) + \,sb(\underline{x}-e_t) \big) 
=  - b(\underline x)  \: .
\end{split}
\end{align}
Choosing $\u = (0,u) \in \Jtest$ with arbitrary $u$,~\eqref{ExGreensCond} yields
\begin{align}
\label{ExGreensCond3}
\begin{split}
\sum_{\underline y \in \UL} sv^\varphi\!(\underline y)  \, f(\underline x - \underline y) = v^\varphi\!(\underline x) \: .
\end{split}
\end{align}
We conclude that in our example, the Green's operator~\eqref{Sdefine} does not mix the scalar
and vector components of $(\Jtest)^\ast$. It may give rise to a constant
component  $(0,S v^0, Sv^1,0) \in \J^\infty$ which, however, drops out of~\eqref{ExWeakEL}
because~$\Delta$ maps constant components of $\J^\infty$ to zero. We use the freedom in the choice 
of Green's operators (cf. Section~\ref{DepGreensOp}) to arrange that~$S v^0 = 0 =Sv^1$.

According to~\eqref{ExGreensCond3}, the vector component
of a Green's operator is a Green's operator of the discrete wave equation
(as usual, one can choose for example the advanced or retarded Green's operator).
The scalar component of the Green's operator, on the other hand,
can be computed by a discrete Fourier transformation.
Since in~\eqref{ExGreensCond2} all functions are evaluated for the same value of~$x^1$, we can solve this
equation for any fixed~$x^1$ and omit this variable.
For the dependence on~$x^0$ we employ the plane-wave ans\"atze
\[ sb(x^0) = a(\omega)\: e^{-i \omega x^0} + \overline{a(\omega)}\: e^{i \omega x^0} \:,\qquad
b(x^0) = b(\omega)\: e^{-i \omega x^0} + \overline{b(\omega)}\: e^{i \omega x^0} \]
with frequency~$\omega \in [0, \pi]$. We thus obtain the equation
\[ a(\omega) = -\frac{b(\omega)}{\lambda_A + 2 \lambda_I\, \cos \omega} \:. \]
In view of~\eqref{lamin}, the denominator is strictly positive, giving a unique solution~$a(\omega)$.
Hence the scalar component of the Green's operator is given by
\beq \label{ScalGreensF}
\hat s(\omega) = -\big(\lambda_A  + 2 \, \lambda_I \: \cos \omega \big)^{-1} \: . 
\eeq
We next evaluate the regularity conditions~(r1) and (r2) on page~\pageref{r1}.
\begin{Lemma} The conditions~\rm{(r1)} and~\rm{(r2)} are satisfied for the
families generated by linearized solutions $\u, \v \in \Jlin$
if $\u, \v$ have vanishing constant vector component (cf. Proposition~\ref{ExLin}),
\beq \label{ExAsmptVan}
\underline{u} = 0 = \underline{v} \: 
\eeq
and if we choose
\beq \label{JtestChoice}
\Jtest_{s,t} = \Jdiff \:,
\eeq
for all $s,t$.
\end{Lemma}
\Proof
Let $\u=(a,u), \, \v=(b,v) \in \Jlin$ with $u$ and $v$ as in~\eqref{ExVecComp} such that~\eqref{ExAsmptVan} holds.
Define $\w^{(1)}_{s,t}$ as in~\eqref{w1def}.
Since, by definition of $S$, its image is in $\J^\infty$, it follows that all non-linear
corrections~$\w_{s,t}^{(2)}, \w_{s,t}^{(3)}, \ldots$ are in $\J^\infty$. Furthermore,
by the above choice of $S$ concerning constant components $(v^0,v^1)$, the
non-linear correction terms have vanishing constant components as well.
It follows that~\eqref{vser}
consists of a scalar component and a vector component in direction of~$v^\varphi$.
Thus the resulting family~\eqref{rhost} consists of functions $f$ and $F$ as described in~\eqref{fFdef},
but where $F$ is constant except for the $S^1$ direction.
Together with the smoothness of the Lagrangian~\eqref{DefL} in the $S^1$-directions $x^\varphi$ and $y^\varphi$, this implies that all partial derivatives in~\eqref{r1Eq} exist.
Moreover, these partial derivatives are $\rho$-integrable
because by the choice of $\rho$ in~\eqref{DefRho} and the bounded support of~$\L(x,.)$,
the $\rho$-integration reduces to a finite sum.
Therefore, the partial derivatives commute with integration, so that~(r1) holds.

The condition~(r2) holds because, by the property of $F$ mentioned in the
last paragraph, the jet~$\u(t)$ as defined in~\eqref{r2Eq} is an element of $\Jdiff$ (given by~\eqref{ExJDiff}) and
hence, due to the choice~\eqref{JtestChoice}, it is also an element of $\Jtest_{0,t} $.
\QED

\subsection{The Conserved Quantities}
We are now in a position to study the corollaries to our main theorem discussed in Section~\ref{SecExamples}.
\subsubsection{The Functional $I^\Omega_1$}
We consider Corollary~\ref{ex0}. Thus let $\v \in \Jlin \cap \Jtest$ have vanishing scalar component, i.e.\
$\v = (0,v)$ with $v$ as in~\eqref{ExVecComp} 
and $\underline{v} = 0$.
The corresponding derivative reads
\begin{align*}
&D_{1,v} \: \L(x,y) \big|_{x^\varphi = 0 = y^\varphi} = - D_{2,v} \: \L(x,y) 
= v^\varphi(\underline x) \frac{d}{dx^\varphi} 
\Big( V(x^\varphi -y^\varphi) \: f(\underline x - \underline y)  \\
& \qquad  \qquad + \delta \, \chi_{B_\varepsilon(0,0)}(\underline{x}-\underline{y})\: V(x^\varphi -y^\varphi )^2 \Big)\big|_{x^\varphi = 0 = y^\varphi} = 0 \: .
\end{align*}
Thus $ I_1^\Omega(\v)$, as defined in~\eqref{I0}, vanishes identically for any $\Omega \subset M$.

\subsubsection{The Functional $I^\Omega_2$ Anti-Symmetrized}
Next, we consider Corollary~\ref{exsymp}.
Define
\begin{align*}
N_t &= \big\{ \underline x \in \UL \,\big|\,  x^0 = t\big\} \quad \text{with} \quad t \in \Z
\end{align*}
and let~$\Omega_{N_t}$ be the past of~$N_t$, i.e.
\beq \label{ChoiceOmega}
\Omega_{N_t} = \big\{ \underline x \in \UL \,\big|\, x^0 \leq t \big\} \:. 
\eeq
As noted in before,~\eqref{I1omegaCons} is exactly the symplectic form $\sigma_\Omega(\u,\v)$ derived in~\cite{jet}.
Thus we have:
\begin{Prp}\label{ExSympPrp} \cite[Proposition~5.8]{jet}
If we choose~$\Jtest$ as the jets with spacelike compact support,
\[ \Jtest = \big\{ \u \in \Jdiff \, \big| \, \text{$\supp \u|_{N_t} $ is a finite set for all~$t \in \Z$} \big\} \:, \]
the conserved functional~\eqref{I1omegaCons} is given by
\begin{align}\label{ExSympl}
\begin{split}
\sigma_{\Omega_{N_t}}(\u, \v) &= \lambda_I  \sum_{\underline x \in N_t} \, \big( a(\underline x) \:b(
\underline{x} + e_t) - a(\underline{x} + e_t) \:b(\underline{x}) \big) \\
& \quad \, \, + \sum_{\underline x \in N_t} \big( \,u^\varphi\!( \underline x+e_t) \,v^\varphi\!(\underline x)   - u^\varphi\!( \underline x) \,v^\varphi\!(\underline{x}+e_t) \big) \:,
\end{split}
\end{align}
where again $e_t := (1,0) \in \R^{1,1}$.
\end{Prp} 
Note that the second sum~\eqref{ExSympl} is the usual symplectic form associated to
a discrete version of the wave equation on $\R^{1,1}$, cf. Remark~5.9 in~\cite{jet}.

\subsubsection{The Functional $I^\Omega_2$ Symmetrized}
We finally compute the surface layer integral of Corollary~\ref{exsprod} and Theorem~\ref{thmintro}.
We first determine $\Delta_2[\u, \v]$. For simplicity, we restrict attention to
the case~\eqref{ScalarVanishAsmpt} of vanishing scalar components.

\begin{Prp} \label{ExPDelta2}
If $\u=(0,u), \v=(0,v) \in \Jlin$, then
the scalar component of $\Delta_2[\u, \v]$ is given by the function 
\beq\label{ExDelta2Eq}
\Delta_2[\u, \v]  (x) = \frac{1}{2} \,  \sum_{\underline y \in \UL}  
 v^\varphi(\underline y) u^\varphi(\underline y) f(\underline x - \underline y)  \:,
 \eeq
whereas the vector component vanishes.
\end{Prp}

\Proof Equation~\eqref{Boxl} gives
\begin{align*}
\Delta_2[\u, \v] (x) &= \frac{1}{2} 
\bigg( \sum_{y \in M} \big( \nabla_{1, \v} + \nabla_{2, \v} \big) \big( \nabla_{1, \u} + \nabla_{2, \u} \big) \, \L(x,y)\:  
-\frac{\nu}{2}\: a(x) \, b(x) \bigg)\\
&= \frac{1}{2}  \sum_{y \in M} \big( D_{1,v} + D_{2, v} \big) \big( D_{1, u} + D_{2, u} \big) \, \L(x,y) \\
&= \frac{1}{2}  \sum_{y \in M} \Big( v^\varphi(\underline x) \frac{d}{dx^\varphi} + v^\varphi(\underline y) \frac{d}{dy^\varphi} \Big)
\Big( u^\varphi(\underline x) \frac{d}{dx^\varphi} + u^\varphi(\underline y) \frac{d}{dy^\varphi} \Big) \, \L(x,y)\\
&= \frac{1}{2}  \sum_{y \in M} \Big( v^\varphi(\underline x)  u^\varphi(\underline x)   \frac{d^2}{d(x^\varphi)^2}
+ \big( v^\varphi(\underline x)\, u^\varphi(\underline y) +  v^\varphi(\underline y) \,u^\varphi(\underline x)  
\big)  \frac{d}{dx^\varphi} \frac{d}{dy^\varphi} \\
&\qquad\qquad \quad+ v^\varphi(\underline y) u^\varphi(\underline y)\frac{d^2}{d(y^\varphi)^2} \Big) \L(x,y) \:,
\end{align*}
where we again used the conventions introduced in Section~\ref{LinSol}.
The second derivative of the Lagrangian~\eqref{DefL} reads
\begin{align}\label{SecondDerivLagr}\begin{split}
 \frac{d^2}{d(x^\varphi)^2} \, \L(x,y) &= f(\underline x - \underline y) \: \cos(x^\varphi-y^\varphi) +  \delta \, \chi_{B_\varepsilon(0,0)}(\underline{x}-\underline{y}) \\
&\qquad \cdot  \big( 2 \, \sin(x^\varphi-y^\varphi)^2 + 2\, (1-\cos(x^\varphi-y^\varphi)) \cos(x^\varphi-y^\varphi) \big) \:.
\end{split}
\end{align}
Evaluated for $x,y \in M$, this gives
\[ 
\frac{d^2}{d(x^\varphi)^2} \, \L(x,y)\big|_{x^\varphi = 0 = y^\varphi} = f(\underline x - \underline y)  \:. \]
For $x \in M$, we thus have
\begin{align*}
\Delta_2[\u, \v] (x) &= \frac{1}{2}  \sum_{\underline y \in \UL}  
\Big(  v^\varphi(\underline x)  u^\varphi(\underline x)  -  v^\varphi(\underline x)u^\varphi(\underline y)  -  v^\varphi(\underline y) u^\varphi(\underline x)  
+ v^\varphi(\underline y) u^\varphi(\underline y) \Big)
 f(\underline x - \underline y) \notag \\
&= \frac{1}{2}  \sum_{\underline y \in \UL}  
\Big(  v^\varphi(\underline y) u^\varphi(\underline y)  -  v^\varphi(\underline x)u^\varphi(\underline y)  -  v^\varphi(\underline y) u^\varphi(\underline x)   \Big)
f(\underline x - \underline y) \:, 
\end{align*}
where in the second step we have used that
$$
\sum_{\underline y \in \UL}  f(\underline x - \underline y) = 0 \: .
$$
According to Proposition~\ref{ExLin}, $v^\varphi$ and $u^\varphi$ satisfy the discrete wave equation~\eqref{ExDiscrWaveEq}.
Thus the last two terms in~\eqref{ExDelta2} vanish, giving
\beq \label{ExDelta2}
\Delta_2[\u, \v] (x) = \frac{1}{2} \,  \sum_{\underline y \in \UL}  
 v^\varphi(\underline y) u^\varphi(\underline y) f(\underline x - \underline y) \:.
\eeq
According to~\eqref{Pairing}, for $\w = (1,0) \in \Jtest$, the scalar component is given by
\[  \nabla_\w \Delta_2[\u, \v] (x) =\Delta_2[\u, \v]  (x)\:, \]
proving the first claim of the proposition.
In order to determine the vector component, we evaluate~\eqref{SecondDerivLagr} for $y \in M$ but $x$ arbitrary,
\begin{align*}
\frac{d^2}{d(x^\varphi)^2} \,  \L(x,y)\big|_{0 = y^\varphi}  &=  f(\underline x - \underline y) \: \cos(x^\varphi) +  \delta \, \chi_{B_\varepsilon(0,0)}(\underline{x}-\underline{y}) \\
&\qquad \cdot  \big( 2 \, \sin(x^\varphi)^2 + 2\, (1-\cos(x^\varphi)) \cos(x^\varphi) \big) \:,
\end{align*}
which implies that
\begin{align*}
&\frac{d^3}{d(x^\varphi)^3} \,  \L(x,y)\big|_{0 = y^\varphi}  = -  f(\underline x - \underline y) \: \sin(x^\varphi) +  \delta \, \chi_{B_\varepsilon(0,0)}(\underline{x}-\underline{y}) \\
&\qquad \times \big( 4 \, \sin(x^\varphi) \cos(x^\varphi)  + 2 \, \sin(x^\varphi) \cos(x^\varphi) - 2\, (1-\cos(x^\varphi)) \sin(x^\varphi) \big) 
\end{align*}
and
\begin{align}\label{ThirdDerivM}
\frac{d^3}{d(x^\varphi)^3} \, \L(x,y)\big|_{x^\varphi=0 = y^\varphi}  &= 0 \:.
\end{align}
Since $\Delta_2$ takes values in $(\Jtest)^\ast$, its vector component merely has an $S^1$-component,
which we denote by~$(\Delta_2[\u, \v])^\varphi(\underline x) $. It is given by
\begin{align*}
&(\Delta_2[\u, \v])^\varphi(\underline x) =  \frac{d}{dx^\varphi} (\Delta_2[\u, \v] ) \\
&=  \frac{d}{dx^\varphi} \frac{1}{2}  \sum_{y \in M} \bigg( v^\varphi(\underline x)  u^\varphi(\underline x)   \frac{d^2}{d(x^\varphi)^2} 
+ ( v^\varphi(\underline x)u^\varphi(\underline y)  \\
&\qquad\qquad +  v^\varphi(\underline y) u^\varphi(\underline x)  )  \frac{d}{dx^\varphi} \frac{d}{dy^\varphi} + v^\varphi(\underline y) u^\varphi(\underline y)\frac{d^2}{d(y^\varphi)^2} \bigg) \L(x,y) \:.
\end{align*}
By our conventions, all derivatives act only on $\L(x,y)$. But~\eqref{ThirdDerivM}
implies that the resulting third derivatives of $\L(x,y)$ vanish on $M$. It follows that
$$(\Delta_2[\u, \v])^\varphi(x) = 0 \qquad \text{ for }  x \in M \: ,$$
proving the second claim of the proposition.
\QED

\noindent In the following, we abbreviate the scalar Green's operator~\eqref{ScalGreensF} applied to the
scalar component of $\Delta_2[\u,\v]$ as given in~\eqref{ExDelta2Eq} by
\[ s(x) := s\,\Delta_2[\u, \v]  (x) \:. \]

\begin{Prp} \label{prp48}
Assume that jets~$\u, \v \in \Jlin \cap \Jtest$ have vanishing scalar components, that the conditions~\eqref{ExAsmptVan}
and~\eqref{JtestChoice} hold, and that the regularity assumptions~{\rm{(r1)}} and~{\rm{(r2)}} 
on page~\pageref{r1} are satisfied.
Then, choosing $\Omega$ as in~\eqref{ChoiceOmega}, the bilinear form~$(.,.)_\Omega$
of Theorem~\ref{thmintro} is given by
\begin{align}\begin{split}\label{ExSym}
(\u,\v)_{\Omega_{N_t}} &= \sum_{\underline x \in N_{t+1}} \big(  u^\varphi(\underline x) \, v^\varphi(\underline x) - 2 \,  \lambda_I   \, s(x) \big)\\
&\quad - \sum_{\underline x \in N_{t}} \big(  u^\varphi(\underline x) \, v^\varphi(\underline x) - 2 \,  \lambda_I   \, s(x) \big)
\end{split}
\end{align}
and satisfies the relation 
\beq \label{ExRel}
(\u,\v)_{\Omega_{N_t}}  = \nu \sum_{x \in \Omega_{N_t}} s(x) \:.
\eeq
\end{Prp}

\Proof
According to Section~\ref{ExGreens}, the Green's operators do not mix the scalar and vector components.
Proposition~\ref{ExPDelta2} implies that the vector component of $S \Delta_2[ \u, \v] $ vanishes,
giving for the bilinear form~\eqref{osiform}
\begin{align*}
(\u,\v)_\Omega &= \int_\Omega d\rho(x) \int_{M \setminus \Omega} d\rho(y) \;
\big( D_{1, u} D_{1, v}   - D_{2, u} D_{2, v} 
+ 2 s(x) - 2 s(y) \big) \, \L(x,y) \\
&= \sum_{x \in \Omega} \ \sum_{y \in M \setminus \Omega} \ \Big(   u^\varphi(\underline x) \, v^\varphi(\underline x) \, 
\partial_{x^\varphi}^2  -  u^\varphi(\underline y) \, v^\varphi(\underline y) \,\partial_{y^\varphi}^2   + 2 s(x) - 2 s(y)  \Big) \L(x,y) \\
& = \sum_{x \in \Omega} \  \sum_{y \in M \setminus \Omega} \ \Big(   u^\varphi(\underline x) \, v^\varphi(\underline x) 
-  u^\varphi(\underline y) \, v^\varphi(\underline y) \Big) f(\underline x - \underline y) \\
& \qquad + 2 \sum_{x \in \Omega} \ \sum_{y \in M \setminus \Omega}  \big(s(x) -  s(y)\big)  \L(x,y) \:, 
\end{align*}
where we again used our conventions for jet derivatives.
Choosing again $\Omega = \Omega_{N_t}$ as in~\eqref{ChoiceOmega}, the second to last term gives
\begin{align*}
&\sum_{\underline x \in N_t} \ \ \sum_{\underline y \in N_{t+1}} \ \Big(   u^\varphi(\underline x) \, v^\varphi(\underline x) 
-  u^\varphi(\underline y) \, v^\varphi(\underline y) \Big) f(\underline x - \underline y) \\
&= \sum_{\underline x \in N_{t+1}} u^\varphi(\underline x) \, v^\varphi(\underline x)  - \sum_{\underline x \in N_t} u^\varphi(\underline x) \, v^\varphi(\underline x) 
\:,
\end{align*}
whereas the last term gives
\begin{align*}
&2 \sum_{x \in \Omega_{N_t}} \ \sum_{y \in M \setminus \Omega_{N_t}} \big( \lambda_A \ \chi_{A}( \underline x - \underline y) + \lambda_I \ \chi_{I}( \underline x - \underline y) \big) \\
&=2 \,  \lambda_I  \sum_{x \in \Omega_{N_t}} \  \sum_{y \in M \setminus \Omega_{N_t}}  \big(s(x) -  s(y)\big)\ \chi_{I}( \underline x - \underline y) \\
&=2 \,  \lambda_I  \Big( \sum_{x \in {N_t}}  s(x) - \sum_{x \in {N_{t+1}}}  s(x)  \Big) \: .
\end{align*}
We thus obtain~\eqref{ExSym}.
For general compact $\Omega$, the right hand side of~\eqref{osirel} is given by
\begin{align*}
\int_\Omega  s(x) \:  \nu  \: d\rho(x) = \nu \sum_{x \in \Omega} s(x) \:,
\end{align*}
giving~\eqref{ExRel}.
\QED

\section{Physical Applications and Outlook} \label{secoutlook}
We now briefly explain the physical significance of the obtained surface layer integrals
and give an outlook on the applications.
From the class of conservation laws found in Theorem~\ref{thmosirho},
the surface layer integrals~$I^\Omega_1$ and~$I^\Omega_2$ have a direct
physical significance. Indeed, a special case of the surface layer integral~$I^\Omega_1$
appears in the Noether-like theorems~\cite{noether}. More generally,
the fact that~\eqref{I1osi} vanishes gives a connection between
a surface layer integral involving the jet~$\u$ and the
space-time integral of the scalar component of~$\u$.
One application we have in mind uses the analogy to general relativity
where the ADM mass, giving the total mass content of the universe,
is defined by a surface integral near spatial infinity. In the limiting case of
Newtonian gravity, this surface integral coincides with the spatial integral
of the mass density. Also in Witten's proof of the positive mass theorem,
the ADM mass is expressed in terms of a spatial integral, in this case
of derivatives of the Witten spinor.
In view of these analogies, it seems an interesting project to explore whether the conservation law
for~$I^\Omega_1$ gives rise to a concept of total mass of a
causal fermion system. As a first step, we plan to analyze static causal fermion systems~\cite{pmt}.

The surface layer integral~$I^\Omega_2$ gives rise to both the symplectic
form~$\sigma_\Omega$ in~\eqref{I1omegaCons} (as introduced earlier in~\cite{jet})
and the surface layer inner product~$(.,.)_\Omega$ in~\eqref{osiform}.
In~\cite{action} these surface layer integrals were computed in detail for Dirac sea
configurations in Minkowski space, with~$\Omega$ chosen as the past of
a Cauchy surface~$t=\text{const}$.
The key result is that for jets describing an electromagnetic
field, the symplectic form agrees with the symplectic form of classical field theory,
whereas the surface layer inner product is positive definite and gives the scalar product
obtained by frequency splitting as used in the usual field quantization procedure.
For jets describing Dirac wave functions, on the other hand, the surface layer inner product is again positive
definite, whereas the symplectic form reproduces the complex structure of the Hilbert space of Dirac solutions.
 In view of these results, the surface layer~$I^\Omega_2$
gives rise to a {\em{scalar product}} on the jet space of linearized solutions
and endows this space with the complex structure used in quantum theory.

In~\cite{fockbosonic}, the connection to bosonic quantum field theory is worked out
in the setting of causal variational principles, again based on the surface layer integral~$I^\Omega_2$.
In~\cite[Section~3.3]{fockbosonic} it is shown with a rescaling technique that the volume term in~\eqref{osirel}
can be arranged to vanish. We thus obtain a scalar product~$(.,.)$
on the linearized solutions which is time independent.
Since the volume term of~$I^\Omega_2$ vanishes when anti-symmetrizing,
the symplectic form is also time independent. We thus obtain two conserved
bilinear forms: the scalar product~$(.,.)$ and the symplectic form~$\sigma(.,.)$.
As shown in~\cite[Section~3]{fockbosonic}, combining these two bilinear forms
gives rise to a {\em{complex structure}} on the linearized solutions.
Describing the nonlinear interaction perturbatively (as developed in~\cite{perturb}),
this complex structure also induces a complex structure on a corresponding Fock space~$\Fock$,
and the dynamics can be described by a linear time evolution on~$\Fock^* \otimes \Fock$
(see~\cite[Sections~4 and~5.2]{fockbosonic}). The usual unitary time evolution on a complex Fock space
is obtained in the so-called holomorphic approximation (see~\cite[Section~5.3]{fockbosonic}).

Since all these constructions make essential use of the surface layer integral~$I^\Omega_2$,
it is fair say that the results of the present paper explain why the dynamics of 
critical points of causal variational principles can be described by a linear time evolution on complex Fock spaces.
We thus obtain a close relation to and a surprising generalizations of concepts of quantum field theory.

The next step in working out the connection to quantum field theory is to include fermionic fields~\cite{fockfermionic}.
Clearly, this cannot be done in the setting of causal variational principles, but one must consider
instead the causal action principle for causal fermion systems~\cite{cfs}.
Again, the resulting constructions will be based on the conservation law for~$I^\Omega_2$
established here.

\appendix
\section{Combinatorics of the Perturbation Expansion} \label{appcombi}
We now give the proof of Lemma~\ref{prpcombi}. In order to facilitate reading,
we give the details of the combinatorics.
We note that a more compact proof, where the perturbation series is rewritten with exponentials,
is given in~\cite[Proof of Lemma~4.1]{perturb}.
\Proof[Proof of Lemma~\ref{prpcombi}]
Our task is to
expand the functional $I_m^\Omega$ in Theorem~\ref{thmosirho} in powers of~$\lambda$.
We first note that~\eqref{cstdef} implies that
\begin{align}\begin{split}\label{f-nach-c}
f^{(p)} &= \frac{1}{p!}\: \frac{d^p}{d\lambda^p} e^{c(\lambda)} \Big|_{\lambda=0} 
=  \frac{1}{p!}\: \frac{d^p}{d\lambda^p} \sum_{\ell=0}^\infty \frac{1}{\ell!} \:(c(\lambda))^\ell \Big|_{\lambda=0}  \\
&= \frac{1}{p!}\: 
\sum_{\ell=1}^p \frac{1}{\ell!} \sum_{\stackrel{\text{\scriptsize{
$q_1, \ldots, q_\ell \geq 1$}}}{\text{with } q_1+\cdots+q_\ell=p}}
\begin{pmatrix} p \\ q_1 \:\cdots\:q_\ell \end{pmatrix}\;
q_1!\:c^{(q_1)} \;\cdots\; q_\ell!\: c^{(q_\ell)}  \\
& = \sum_{\ell=1}^p \frac{1}{\ell!} \sum_{\stackrel{\text{\scriptsize{
$q_1, \ldots, q_\ell \geq 1$}}}{\text{with } q_1+\cdots+q_\ell=p}}
c^{(q_1)} \:\cdots\: c^{(q_\ell)} \:,
\end{split}
\end{align}
where in the second step we used that~$c^{(0)} = 0$ 
in order to truncate the summation over $l$ as well as the general Leibniz rule for differentiable functions.

We next perform the perturbation expansion of the Lagrangian. To this end, we first expand it in a Taylor series in both arguments. In the following formula,
$(D_{1, F(x)-x}) \L(x,y)$ and~$(D_{2, F(y)-y}) \L(x,y)$ again denote the partial derivatives of~$\L(.,.)$ acting on the first and second argument,
in direction of $F(x)-x$ and $F(y)-y$, respectively.
According to our general convention~\eqref{ConventionPartial}, these partial derivatives do not act on the arguments~$F(x)-x$ or~$F(y)-y$ of other derivative, but merely on~$\L$. We thus obtain
\begin{align*}
\L\big(F(x), F(y) \big) &= \sum_{a=0}^\infty \frac{1}{a!} \sum_{b=0}^\infty \frac{1}{b!}\;
\big(D_{1, F(x)-x}\big)^a \big(D_{2, F(y)-y}\big)^b \L(x,y) \\
&= \sum_{k=0}^\infty \sum_{\stackrel{\text{\scriptsize{$a,b\geq0$}}}{a+b = k}} \frac{1}{a! \:b!}\;
\big(D_{1, F(x)-x}\big)^a \big(D_{2, F(y)-y}\big)^b \L(x,y) \\
&= \sum_{k=0}^\infty \frac{1}{k!} \:\Big(D_{1, F(x)-x} + D_{2, F(y)-y} \Big)^k \L(x,y) \:,
\end{align*}
where the second step consists merely of a reordering of summation indices, and where
 in the last step we used the generalized binomial theorem.
Taking the $q^\text{th}$ derivative with respect to~$\lambda$ and evaluating at~$\lambda=0$,
the general Leibniz rule yields
\begin{align}\begin{split}\label{DerivativeL}
&\frac{1}{q!} \frac{d^q}{d\lambda^q}\: \L\big(F(x), F(y) \big) \Big|_{\lambda=0}
= \frac{1}{q!} \:\sum_{k=0}^q \frac{1}{k!}\;\;\sum_{\stackrel{\text{\scriptsize{$q_1, \ldots, q_k \geq 1$}}}
{\text{with } q_1+\cdots+q_k=q}} \begin{pmatrix} q \\ q_1 \:\cdots\: q_k \end{pmatrix} \notag \\
&\qquad \times\:q_1!\, \big(D_{1, F^{(q_1)}} + D_{2, F^{(q_1)}} \big) \cdots 
q_k!\, \big(D_{1, F^{(q_k)}} + D_{2, F^{(q_k)}} \big)  \L(x,y)  \\
&= \sum_{k=0}^q \frac{1}{k!}\;\;\sum_{\stackrel{\text{\scriptsize{$q_1, \ldots, q_k \geq 1$}}}
{\text{with } q_1+\cdots+q_k=q}} \:
\big(D_{1, F^{(q_1)}} + D_{2, F^{(q_1)}} \big) \cdots \big(D_{1, F^{(q_k)}} + D_{2, F^{(q_k)}} \big)  \L(x,y) \:,
\end{split}
\end{align}
where in the first step we used the equation~$(D_{1, F^{(0)}(x)-x}) \L(x,y) = 0$
in order to truncate the summation over $k$. Moreover, differentiating the expansion~\eqref{Fxser}
gives
\[  \frac{d^{q_i}}{d\lambda^{q_i}}\: \big(D_{1, F(x)-x} + D_{2, F(y)-y}\big) = q_i! \: \big(D_{1, F^{(q_i)}} + D_{2, F^{(q_i)}}
\big) \:. \]
We remark that the term for~$k=0$ only contributes if $q=0$.
Using again the general Leibniz rule, we have
\begin{align*} 
&\frac{1}{p!} \frac{d^p}{d\lambda^p} \: f(x) \:\L\big(F(x), F(y) \big)\: f(y)\:\\
&= \sum_{\stackrel{\text{\scriptsize{$l,r,q \geq 0$}}} {\text{with } l+r+q=p}}
f^{(l)}(x)\:f^{(r)}(y) \:\frac{1}{q!} \frac{d^q}{d\lambda^q}\L\big(F(x), F(y) \big)\Big|_{\lambda=0} \:.
\end{align*}
The factors $f^{(l)}(x)\:f^{(r)}(y)$ can be calculated similar to~\eqref{f-nach-c},
\begin{align*}
&\sum_{\stackrel{\text{\scriptsize{$l,r \geq 0$}}}{\text{with } l+r=L}}
\!\!\!\!f^{(l)}(x)\: f^{(r)}(y)
= \sum_{\stackrel{\text{\scriptsize{$l,r \geq 0$}}}{\text{with } l+r=L}}\!\!\!\!
\frac{1}{l!}\: \frac{d^l}{d\lambda^l} e^{c(x)}\; \frac{1}{r!}\: \frac{d^r}{d\lambda^r} e^{c(y)} \Big|_{\lambda=0} \\
&=\frac{1}{L!} \frac{d^L}{d\lambda^L} e^{c(x) + c(y)} \Big|_{\lambda=0} = \frac{1}{L!} \frac{d^L}{d\lambda^L} \sum_{a=0}^\infty \frac{1}{a!} (c(x) + c(y))^a \Big|_{\lambda=0} \\
&= \frac{1}{L!}\: \sum_{a=0}^L \frac{1}{a!} \!\!\!\!
\sum_{\stackrel{\text{\scriptsize{$l_1, \ldots, l_a \geq 1$}}}{\text{with } l_1+\cdots+l_a=L}} \!\!\!\!\!\!
\begin{pmatrix} L \\ l_1 \:\cdots\:l_a \end{pmatrix}  l_1!\:\big( c^{(l_1)}(x) +  c^{(l_1)}(y) \big)
\; \cdots \; l_a!\: \big(c^{(l_a)}(x) + c^{(l_a)}(y) \big) \\
&\quad = \sum_{a=0}^L \frac{1}{a!} \!\!\!\!
\sum_{\stackrel{\text{\scriptsize{$l_1, \ldots, l_a \geq 1$}}}{\text{with } l_1+\cdots+l_a=L}}  \!\!\!\!
\big( c^{(l_1)}(x) +  c^{(l_1)}(y) \big) \;\cdots \;\big(c^{(l_a)}(x) + c^{(l_a)}(y) \big) \:.
\end{align*}
\nopagebreak
Combing the above formulas, we obtain
\begin{align}
&\frac{1}{p!} \frac{d^p}{d\lambda^p} \: f(x) \:\L\big(F(x), F(y) \big)\: f(y)\:\notag \\
&= \sum_{\stackrel{\text{\scriptsize{$l,r,q \geq 0$}}} {\text{with } l+r+q=p}} \sum_{k=0}^q \frac{1}{k!}\;\;\sum_{\stackrel{\text{\scriptsize{$q_1, \ldots, q_k \geq 1$}}} {\text{with } q_1+\cdots+q_k=q}} \notag \\
&\qquad \times f^{(l)}(x)\:f^{(r)}(y) \big(D_{1, F^{(q_1)}} + D_{2, F^{(q_1)}} \big) \cdots \big(D_{1, F^{(q_k)}} + D_{2, F^{(q_k)}} \big)  \L(x,y) \notag \\
&= \sum_{q=0}^p \sum_{\stackrel{\text{\scriptsize{$l,r \geq 0$}}} {\text{with } l+r=p-q}} \sum_{k=0}^q \frac{1}{k!}\;\;\sum_{\stackrel{\text{\scriptsize{$q_1, \ldots, q_k \geq 1$}}} {\text{with } q_1+\cdots+q_k=q}} \notag \\
&\qquad \times f^{(l)}(x)\:f^{(r)}(y) \big(D_{1, F^{(q_1)}} + D_{2, F^{(q_1)}} \big) \cdots \big(D_{1, F^{(q_k)}} + D_{2, F^{(q_k)}} \big)  \L(x,y) \notag \\
&= \sum_{q=0}^p \sum_{a=0}^{p-q} \frac{1}{a!}   \sum_{k=0}^q \frac{1}{k!}\;\;\sum_{\stackrel{\text{\scriptsize{$q_1, \ldots, q_k \geq 1$}}} {\text{with } q_1+\cdots+q_k=q}}  \ \ 
\sum_{\stackrel{\text{\scriptsize{$l_1, \ldots, l_a \geq 1$}}}{\text{with } l_1+\cdots+l_a=p-q}}  \notag \\
&\qquad \times \big( c^{(l_1)}(x) +  c^{(l_1)}(y) \big) \;\cdots \;\big(c^{(l_a)}(x) + c^{(l_a)}(y) \big) \notag \\
&\qquad \times \big(D_{1, F^{(q_1)}} + D_{2, F^{(q_1)}} \big) \cdots \big(D_{1, F^{(q_k)}} + D_{2, F^{(q_k)}} \big)  \L(x,y) \notag \\
&= \sum_{l=0}^p \sum_{\stackrel{\text{\scriptsize{$a,k \geq 0$}}} {a+k=l}} \frac{1}{a!\, k!}\;\; \sum_{\stackrel{\text{\scriptsize{$l_1, ..., l_a , q_1, \ldots, q_k \geq 1$}}} {\text{with } l_1 + ... + l_a + q_1+\cdots+q_k=p}}  \notag \\
&\qquad \times \big( c^{(l_1)}(x) +  c^{(l_1)}(y) \big) \;\cdots \;\big(c^{(l_a)}(x) + c^{(l_a)}(y) \big) \notag \\
&\qquad \times \big(D_{1, F^{(q_1)}} + D_{2, F^{(q_1)}} \big) \cdots \big(D_{1, F^{(q_k)}} + D_{2, F^{(q_k)}} \big)  \L(x,y) \: , \label{calcN}
\end{align}
where in the last step we argued as follows. 
Note that for a given value of~$(a,k)$, in the second-to-last line, the sum over $q$ gives~$(p-(a+k))$ terms since $\ell_i,\,q_i \geq 1$. The sum
over these $(p-(a+k))$ terms reads
\begin{align*}
&\sum_{q=k}^{p-a} \sum_{\stackrel{\text{\scriptsize{$q_1, \ldots, q_k \geq 1$}}} {\text{with } q_1+\cdots+q_k=q}}  \ \ 
\sum_{\stackrel{\text{\scriptsize{$l_1, \ldots, l_a \geq 1$}}}{\text{with } l_1+\cdots+l_a=p-q}}
\cdots \\
&= \sum_{q=0}^{p} \sum_{\stackrel{\text{\scriptsize{$q_1, \ldots, q_k \geq 1$}}} {\text{with } q_1+\cdots+q_k=q}}  \ \ 
\sum_{\stackrel{\text{\scriptsize{$l_1, \ldots, l_a \geq 1$}}}{\text{with } l_1+\cdots+l_a=p-q}} \cdots 
\;=\; \sum_{\stackrel{\text{\scriptsize{$l_1, ..., l_a , q_1, \ldots, q_k \geq 1$}}} {\text{with } l_1 + ... + l_a + q_1+\cdots+q_k=p}}  \cdots \:,
\end{align*}
where in the second step we used that extending the range of $q$ does 
not change the terms which are summed over, because the conditions $\ell_i,\,q_i \geq 1$ exclude those terms for fixed $k$ and $a$.
It remains to sum over the \emph{different} values of $(a,k)$, which is most conveniently done via
\[ 
\sum_{\ell=0}^p \sum_{\stackrel{\text{\scriptsize{$a,k \geq 0$}}} {a+k=\ell}} \cdots \: ,
\]
giving the result of the calculation~\eqref{calcN}.

Next, we use the definition of $\w^{(p)}$ in~\eqref{vser}. We have
\begin{align*}
&\sum_{\stackrel{\text{\scriptsize{$q_1, \ldots, q_\ell \geq 1$}}} {\text{with } q_1+\cdots+q_\ell=p}} \nabla_{1,\w^{(q_1)}} \nabla_{1,\w^{(q_2)}} \cdots \nabla_{1,\w^{(q_\ell)}} \L(x,y)\\
&= \sum_{\stackrel{\text{\scriptsize{$q_1, \ldots, q_l \geq 1$}}} {\text{with } q_1+\cdots+q_\ell=p}} (c^{(q_1)}(x)+D_{1,F^{(q_1)}}) (c^{(q_2)}(x)+D_{1,F^{(q_2)}}) \cdots (c^{(q_\ell)}(x)+D_{1,F^{(q_\ell)}}) \L(x,y)\\
&=  \sum_{\stackrel{\text{\scriptsize{$a,k \geq 0$}}} {a+k =\ell}}  \begin{pmatrix} \ell \\ a \ k \end{pmatrix} \! \!
 \sum_{\stackrel{\text{\scriptsize{$l_1, ..., l_a , q_1, \ldots, q_k \geq 1$}}} {\text{with } l_1 + ... + l_a + q_1+\cdots+q_k=p}} \! \!
c^{(l_1)}(x) \cdots c^{(l_a)}(x) \: D_{1,F^{(q_1)}} \cdots D_{1,F^{(q_k)}}  \L(x,y)	\:.
\end{align*}
The last step can be understood as follows. In order to get from the second line to the third line, from every bracket $(c^{(q_i)}(x)+D_{1,F^{(q_i)}})$ we choose either $c^{(q_i)}(x)$ or $D_{1,F^{(q_i)}}$. 
The order of the appearance of the derivatives does not matter because, as explained after~\eqref{f-nach-c}, the derivatives all act on the first argument of $\L(x,y)$ and not on each other nor on the $c^{(q_i)}(x)$. Furthermore, the actual value of $q_i$ does not matter because we sum over all $q_i$. Thus every term in the second line is characterized by how many $c^{(q_i)}(x)$ appear (denote this number be $a$) and by how many $D_{1,F^{(q_i)}}$ appear (we denote this number be $k$), giving rise to the sum over $a$ and $k$ in the last line. The binomial coefficient gives the correct combinatorial factor, corresponding to the number of ways
that, disregarding ordering, we can choose~$a$ (respectively $k$) terms from $l$ terms.

The same argument as in the last paragraph can be applied to terms $(\nabla_{1,\w^{(q_i)}} + \nabla_{2,\w^{(q_i)}})$ instead of $\nabla_{1,\w^{(q_i)}}$,
giving exactly the terms appearing in the last term in~\eqref{calcN}. Thus the
$p^\text{th}$ order in the perturbation expansion of the integrand of~\eqref{Ikdef0} can be expressed as
\begin{align}\begin{split}\label{kompl}
&\frac{1}{p!} \frac{d^p}{d\lambda^p} \: f(x) \:\L\big(F(x), F(y) \big)\: f(y)\:\\
&=\sum_{\ell=0}^p \frac{1}{\ell !} \! \! \sum_{\stackrel{\text{\scriptsize{$q_1, \ldots, q_\ell \geq 1$}}} {\text{with } q_1+\cdots+q_\ell=p}} \!\!\!\! (\nabla_{1,\w^{(q_1)}} + \nabla_{2,\w^{(q_1)}}) \cdots 
(\nabla_{1,\w^{(q_\ell)}} + \nabla_{2,\w^{(q_\ell)}}) \, \L(x,y) \, .
\end{split}\end{align}

The remaining task is to distribute the $s$- and $t$-derivatives in Theorem~\ref{thmosirho}.
Since the term for~$\ell=0$ in~\eqref{kompl} only contributes if $p=0$ and thus, according to~\eqref{fxser} and~\eqref{Fxser},
does not have a dependence on $s$ or $t$, the summation over $\ell$ does not need to include the case $\ell=0$.
We thus obtain
\begin{align*}
& \big(\partial_{1,s} - \partial_{2,s} \big) \big( \partial_{1,t} + \partial_{2,t} \big)^k  \frac{1}{p!} \frac{d^p}{d\lambda^p} \: f_{s,t}(x) \:\L\big(F_{s,t}(x), F_{s,t}(y) \big)\: f_{s,t}(y) \Big|_{s=t=0} \\
&= \sum_{\ell=1}^p \frac{1}{\ell!} \! \! \sum_{\stackrel{\text{\scriptsize{$q_1, \ldots, q_\ell \geq 1$}}} {\text{with } q_1+\cdots+q_\ell=p}} \!\!\!\! \big(\partial_{1,s} - \partial_{2,s} \big) \big( \partial_{1,t} + \partial_{2,t} \big)^k  \\
&\qquad \qquad \times (\nabla_{1,\w_{s,t}^{(q_1)}} + \nabla_{2,\w_{s,t}^{(q_1)}}) \cdots 
(\nabla_{1,\w_{s,t}^{(q_\ell)}} + \nabla_{2,\w_{s,t}^{(q_\ell)}}) \, \L(x,y) \Big|_{s=t=0} \\
&= \sum_{\ell=1}^p \frac{1}{\ell!} \! \! \sum_{\stackrel{\text{\scriptsize{$q_1, \ldots, q_\ell \geq 1$}}} {\text{with } q_1+\cdots+q_\ell=p}} \ \
\sum_{\stackrel{\text{\scriptsize{$k_1, \ldots, k_\ell \geq 0$}}} {\text{with } k_1+\cdots+k_\ell=k}} \!\!\! {{k}\choose{k_1 \, \ldots \, k_\ell}} \: \big(\partial_{1,s} - \partial_{2,s} \big) \\
& \qquad  \times (\nabla_{1,\partial_t^{k_1}\w_{s,t}^{(q_1)}} + \nabla_{2,\partial_t^{k_1}\w_{s,t}^{(q_1)}}) \cdots 
(\nabla_{1,\partial_t^{k_\ell}\w_{s,t}^{(q_\ell)}} + \nabla_{2,\partial_t^{k_\ell}\w_{s,t}^{(q_\ell)}}) \, \L(x,y) \Big|_{s=t=0}\\
&= \sum_{\ell=1}^p \frac{1}{(\ell-1)!} \! \! \sum_{\stackrel{\text{\scriptsize{$q_1, \ldots, q_\ell \geq 1$}}} {\text{with } q_1+\cdots+q_\ell=p}} \ \
\sum_{\stackrel{\text{\scriptsize{$k_1, \ldots, k_\ell \geq 0$}}} {\text{with } k_1+\cdots+k_\ell=k}} \!\!\! {{k}\choose{k_1 \, \ldots \, k_\ell}} \:\\
& \qquad  \times (\nabla_{1,\partial_s\partial_t^{k_1}\w_{s,t}^{(q_1)}} - \nabla_{2,\partial_s\partial_t^{k_1}\w_{s,t}^{(q_1)}}) \cdots 
(\nabla_{1,\partial_t^{k_\ell}\w_{s,t}^{(q_\ell)}} + \nabla_{2,\partial_t^{k_\ell}\w_{s,t}^{(q_\ell)}}) \, \L(x,y) \Big|_{s=t=0} \:.
\end{align*}
Note that the summation over $k_1, ..., k_\ell$ needs to include the cases $k_i = 0$ because
in general~$\w^{(p)}_{s,t} |_{s=0=t} \neq 0$ (see~\eqref{vser}).
For the integrand in~\eqref{Ikdef}, using~\eqref{f-nach-c}, we obtain for the $p^\text{th}$ order
\begin{align*}
&\partial_{s} \partial_{t}^k  f^{(p)}(x) \Big|_{s=t=0} =  \sum_{\ell=1}^p \frac{1}{\ell!} \! \! \sum_{\stackrel{\text{\scriptsize{ $q_1, \ldots, q_\ell \geq 1$}}}{\text{with } q_1+\cdots+q_\ell=p}} \! \! 
\partial_{s} \partial_{t}^k \big( c^{(q_1)} \:\cdots\: c^{(q_\ell)} \big) \Big|_{s=t=0}\\
&=  \sum_{\ell=1}^p \frac{1}{(\ell-1)!} \!\!\!\!\! \sum_{\stackrel{\text{\scriptsize{ $q_1, \ldots, q_\ell \geq 1$}}}{\text{with } q_1+\cdots+q_\ell=p}} \;
\sum_{\stackrel{\text{\scriptsize{ $k_1, \ldots, k_\ell \geq 0$}}}{\text{with } k_1+\cdots+k_\ell=k}} 
\!\!\!\! \begin{pmatrix} k \\ k_1 \ \ldots k_\ell \end{pmatrix} 
\big(\partial_{s}  \partial_t^{k_1} c^{(q_1)}) \:\cdots\: (\partial_t^{k_\ell}  c^{(q_\ell)} \big) \Big|_{s=t=0}\: .
\end{align*}
This gives the result.
\QED

\Thanks {{\em{Acknowledgments:}}
We would like to thank Andreas Platzer and the referee for helpful comments on the manuscript.


\begin{thebibliography}{10}

\bibitem{bogachev}
V.I. Bogachev, \emph{Measure {T}heory. {V}ol. {I}}, Springer-Verlag, Berlin,
  2007.

\bibitem{linhyp}
C.~Dappiaggi and F.~Finster, \emph{Linearized fields for causal variational
  principles: {E}xistence theory and causal structure}, arXiv:1811.10587
  [math-ph] (2018).

\bibitem{fockfermionic}
F.~Finster, \emph{Fermionic {F}ock space dynamics for causal fermion systems},
  in preparation.

\bibitem{continuum}
\bysame, \emph{Causal variational principles on measure spaces},
  arXiv:0811.2666 [math-ph], J. Reine Angew. Math. \textbf{646} (2010),
  141--194.

\bibitem{cfs}
\bysame, \emph{The {C}ontinuum {L}imit of {C}ausal {F}ermion {S}ystems},
  arXiv:1605.04742 [math-ph], Fundamental Theories of Physics, vol. 186,
  Springer, 2016.

\bibitem{action}
\bysame, \emph{The causal action in {M}inkowski space and surface layer
  integrals}, arXiv:1711.07058 [math-ph] (2017).

\bibitem{perturb}
\bysame, \emph{Perturbation theory for critical points of causal variational
  principles}, arXiv:1703.05059 [math-ph] (2017).

\bibitem{fockbosonic}
F.~Finster and N.~Kamran, \emph{Complex structures on jet spaces and bosonic
  {F}ock space dynamics for causal variational principles}, arXiv:1808.03177
  [math-ph] (2018).

\bibitem{dice2014}
F.~Finster and J.~Kleiner, \emph{Causal fermion systems as a candidate for a
  unified physical theory}, arXiv:1502.03587 [math-ph], J. Phys.: Conf. Ser.
  \textbf{626} (2015), 012020.

\bibitem{noether}
\bysame, \emph{Noether-like theorems for causal variational principles},
  arXiv:1506.09076 [math-ph], Calc. Var. Partial Differential Equations
  \textbf{55:35} (2016), no.~2, 41.

\bibitem{jet}
\bysame, \emph{A {H}amiltonian formulation of causal variational principles},
  arXiv:1612.07192 [math-ph], Calc. Var. Partial Differential Equations
  \textbf{56:73} (2017), no.~3, 33.

\bibitem{pmt}
F.~Finster and A.~Platzer, \emph{The total mass of a static causal fermion
  system}, in preparation.

\bibitem{support}
F.~Finster and D.~Schiefeneder, \emph{On the support of minimizers of causal
  variational principles}, arXiv:1012.1589 [math-ph], Arch. Ration. Mech. Anal.
  \textbf{210} (2013), no.~2, 321--364.

\bibitem{halmosmt}
P.R. Halmos, \emph{Measure {T}heory}, Springer, New York, 1974.

\bibitem{rudin}
W.~Rudin, \emph{Real and {C}omplex {A}nalysis}, third ed., McGraw-Hill Book
  Co., New York, 1987.

\end{thebibliography}
\providecommand{\bysame}{\leavevmode\hbox to3em{\hrulefill}\thinspace}
\providecommand{\MR}{\relax\ifhmode\unskip\space\fi MR }
\providecommand{\MRhref}[2]{%
  \href{http://www.ams.org/mathscinet-getitem?mr=#1}{#2}
}
\providecommand{\href}[2]{#2}

\end{document}